\documentclass[journal,draftclsnofoot,onecolumn,12pt]{IEEEtranTCOM}

\usepackage{graphicx}
\usepackage{amssymb}
\usepackage{multirow}
\usepackage{tabularx}
\usepackage{acronym}
\usepackage{color}
\usepackage{lettrine}
\usepackage{mdwlist}
\usepackage{gensymb}
\usepackage{paralist}
\usepackage{cite}
\usepackage[cmex10]{amsmath}
\usepackage[font=footnotesize]{caption}
\usepackage[font=footnotesize]{subcaption}
\usepackage[utf8]{inputenc}
\usepackage{url}

\normalsize

\ifCLASSINFOpdf
\else
\fi

\begin{document}
%
\title{Physical Layer Performance Evaluation of Wireless Infrared-based LiFi Uplink}

\author{Cheng~Chen,~\IEEEmembership{Student Member,~IEEE,}
	Dushyantha~Basnayaka,~\IEEEmembership{Senior Member,~IEEE,}
	Ardimas~Andi~Purwita,~\IEEEmembership{Student Member,~IEEE,}
	Xiping~Wu,~\IEEEmembership{Member,~IEEE,}
	and~Harald~Haas,~\IEEEmembership{Fellow,~IEEE}
	\thanks{The work of Professor H. Haas was supported by EPSRC under Established Career Fellowship Grant EP/R007101/1. The authors are with the Li-Fi Research and Development Centre, Institute for Digital Communications, The University of Edinburgh, EH9 3JL, Edinburgh, UK e-mail: \{cheng.chen, d.basnayaka, a.purwita, xiping.wu, h.haas\}@ed.ac.uk.
	}}


\maketitle

\begin{abstract}
LiFi (light-fidelity) is recognised as a promising technology for the next generation wireless access network. However, little research effort has been spent on the uplink transmission system in LiFi networks. In this paper, an analytical framework for the performance analysis of an wireless infrared-based LiFi uplink at system level is presented. This work focuses on the performance of a single user who is randomly located in a network.  Many important factors in practice are taken into account, such as front-end characteristics, channel path loss, link blockage and user device random orientations. In particular, accurate analytical expressions for the statistics of optical channel path loss has been derived. Based on this path loss statistics, the distribution of signal-to-noise ratio (SNR) and the average achievable data rate are evaluated. A channel factor is defined as a quantity which only depends on the used devices, optical channel and noise power spectral density function (PSD). The result shows that with commercial front-end elements, an average data rate of up to 250~Mbps is achievable.
\end{abstract}

\begin{IEEEkeywords}
	Light-fidelity, wireless infrared communication, visible light communication, link blockage.
\end{IEEEkeywords}

%
\IEEEpeerreviewmaketitle

\section{Introduction}
\IEEEPARstart{W}{ith} the already overcrowded radio frequency (RF) spectrum resources, it becomes more and more challenging to achieve wireless communications that can fulfil the increasing wireless data traffic demand. Consequently, wireless communication based on spectrum resources in high frequency ranges have been considered. Visible light communication is one of the most promising solutions, which is able to establish a link with data rate of more than 10~Gbps \cite{Islim:17}. Based on visible light communication (VLC) technology, a small-cell wireless access network has been proposed, which is known as LiFi (light-fidelity). Compared to VLC, LiFi is a more unique and multifunctional system, which uses multiple coordinated light sources in an indoor environment as access points (APs). Each AP provides connectivity to a few nearby mobile users. A LiFi network is able to provide duplex transmission between an AP and a user equipment (UE). It also supports multiple access, handover and many other powerful functionalities \cite{Haas2016}. Due to the fact that light signal is directional and cannot penetrate opaque objects, LiFi networks can provide more secure connectivity in an indoor environment compared to conventional RF access networks \cite{8119998}. In addition, LiFi networks achieve a considerable increase in data rate per unit area, which makes LiFi advantageous over RF small-cell systems in densely populated scenarios \cite{cbh1601}.

In a number of survey publications, the possible technologies for LiFi uplink have been discussed \cite{7072557,pfhm1501}, where the uplink transmissions using RF, VLC and wireless infrared have been mentioned the most. A comparison between various existing RF technologies that can be used for LiFi uplink has been presented in \cite{7841270}. WiFi or other RF technology has been considered as the uplink system in a RF-optical heterogeneous network \cite{7314097}. However, the use of RF-based uplink introduces interference with existing RF systems. Furthermore, RF technology cannot be used in electromagnetic interference (EMI)-sensitive areas. The use of VLC technology in the LiFi uplink has been considered and experimentally demonstrated in \cite{6935084,6532859}. However, visible light uplink signal may cause interference to the VLC-based LiFi downlink transmission. In addition, it also causes discomfort to the eyes of the mobile users. The propagation of a wireless infrared signal is similar to that of a visible light signal, but the signal spectrum is in the infrared region which is invisible to the human eye. Thus, it causes no discomfort to mobile users. More importantly, it does not interfere with either RF or VLC-based systems. Wireless infrared-based uplink transmission has been studied using computer simulations \cite{Sewaiwar:15,7847572} and demonstrated using proof-of-concept experiments \cite{6783722,6261653}. To enhance the reliability of the wireless infrared link, beam steering and LED allocation schemes have been proposed \cite{7847572,6261653}. Concerns about channel uncertainty caused by random UE position/orientation and link blockage have been raised by many publications, but few studies have conducted a systematic investigation on this issue. In this paper, an analytical framework for the statistics of the wireless infrared-based uplink channel has been presented. In the analysis, the characteristics of used front-end elements, random UE location, random UE orientation and link blockage by human bodies have been taken into account. Based on the channel statistics, the signal-to-noise ratio (SNR) statistics and average achievable data rate are evaluated.

The remainder of this paper is organised as follows: the wireless infrared uplink system model is introduced in Section~\ref{sec:uplink system model}. The uplink optical channel and considered channel uncertainties are introduced in Section~\ref{sec:LoS_characteristics}. The statistics of the uplink optical channel are presented in Section~\ref{sec:channel_path_loss_statistics}. Based on optical channel statistics, the channel factor statistics and achievable data rate are evaluated in Section~\ref{sec:SNR_data_rate}. Finally, conclusions are drawn in Section~\ref{sec:conclusion}. 

The following notations are used throughout this paper: $\{\cdot\}^{\mathrm{T}}$ is the transpose of a vector or matrix; $\mathbb{E}\{\cdot\}$ represents the statistical expectation; $\delta(u)$ is the unit Dirac delta function; $\max\{\cdot\}$ is the maxima operator; $\otimes$ represents the convolution operator; $\mathfrak{u}(u)$ denotes the unit step function; $\mathrm{Ei}[x]=\int_{-\infty}^{x}\frac{1}{t}\exp(t)\mathrm{d}t$ is the exponential integral function; $\mathcal{Q}(u)=\frac{1}{\sqrt{2\pi}}\int_u^{\infty}\exp(-v^2/2)\mathrm{d}v$ is the Q-function;  $\mathfrak{f}_{\mathcal{N}}(u)=\frac{1}{\sqrt{2\pi}}\exp\left(-u^2/2\right)$ is the probability density function of the standard normal distribution; $\mathrm{sgn}(u)$ is the sign function; $[x]^{+}=\max(x,0)$ denotes the ramp function and a similar function is defined as $[x]^{-}=\min(x,0)$.

\begin{figure}
	\centering
	\begin{subfigure}{.47\textwidth}
		\centering
		\includegraphics[width=1\textwidth]{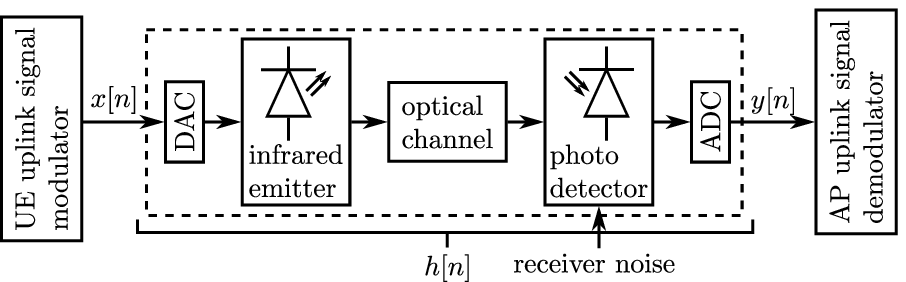}
		\caption{}
		\label{fig:block_uplink_channel}
	\end{subfigure}	
	\begin{subfigure}{.47\textwidth}
		\centering
		\includegraphics[width=1\linewidth]{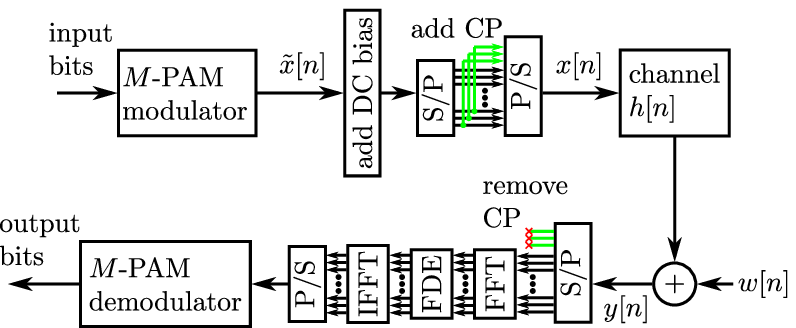}
		\caption{}
		\label{fig:block_pam}
	\end{subfigure}%
	\caption{(a) Wireless infrared-based uplink block diagram. (b) PAM-SCFDE block diagram.}
	\label{fig:comb1}
\end{figure}


\section{Wireless Infrared Uplink System Model}
\label{sec:uplink system model}

\subsection{Wireless Infrared Uplink Channel Model}
The wireless infrared-based uplink system can be described by the block diagram as shown in Fig.~\ref{fig:comb1}~(a). The discrete-time transmitted signal is denoted by $x[n]$, which has variance of $\sigma_{x}^2$. The transmitted signal $x[n]$ is forwarded to a channel comprising a digital-to-analogue converter (DAC), an infrared light emitting diode (LED), an optical channel, a photodiode (PD) and an analogue-to-digital converter (ADC). The time domain characteristics of this channel can be described by an impulse response: $ \mathfrak{h}(t)=\int_{\lambda_{\rm min}}^{\lambda_{\rm max}}\mathfrak{h}_{\lambda}(t)\mathrm{d}\lambda$, where $\lambda$ is the optical signal wavelength, $\lambda_{\rm min}$ ($\lambda_{\rm max}$) is the minimum (maximum) wavelength of interest, and $\mathfrak{h}_{\lambda}(t)$ represents the impulse response of the channel at wavelength $\lambda$. $\mathfrak{h}_{\lambda}(t)$ can be calculated by:
\begin{align}
\mathfrak{h}_{\lambda}(t)=\mathfrak{h}_{\rm dac}(t)\otimes\mathfrak{h}_{{\rm led}}^{\lambda}(t)\otimes\mathfrak{h}_{\rm oc}(t)\otimes\mathfrak{h}_{{\rm pd}}^{\lambda}(t)\otimes\mathfrak{h}_{\rm adc}(t),
\label{eq:analogue_cir}
\end{align} 
where $\mathfrak{h}_{\rm dac}(t)$ and $\mathfrak{h}_{\rm adc}(t)$ represent the impulse response of the DAC and the ADC, which are responsible for the conversion between digital signals and analogue signals. It is assumed that the bandwidth of the DAC and the ADC are sufficiently wide compared to the modulation bandwidth so that the inter-symbol interference (ISI) caused by DAC and ADC is negligible. The impulse response of the LED at wavelength $\lambda$ can be defined as:
\begin{align}
\mathfrak{h}_{{\rm led},\lambda}(t)=\sigma_{\rm o}\mathcal{E}_{\rm led}(\lambda)\mathfrak{h}_{\rm lp,led}(t), 
\label{eq:led_cir}
\end{align}
where $\sigma_{\rm o}$ is a signal scaling factor, which adjust the amplitude of the optical signal to fit the linear dynamic range of the LED. The LED has a maximum optical power of $P_{\rm o,max}$ and a minimum optical power output of $P_{\rm o,min}$. Thus, the linear dynamic range of the LED can be defined as: $\Delta P_{\rm o}=P_{\rm o,max}-P_{\rm o,min}$. A perfect linear conversion between the electrical input and the optical output within this dynamic range is assumed. Considering a signal with a crest factor of $\eta_{\rm c}$\footnote{crest factor is the square root of peak-to-average power ratio (PAPR).}, the signal scaling factor can be calculated as: $\sigma_{\rm o}=\frac{\Delta P_{\rm o}}{2\eta_{\rm c}\sigma_{x}}$. $\mathcal{E}_{\rm led}(\lambda)$ in \eqref{eq:led_cir} is the normalised spectral emission, which shows the proportion of the emitted optical power at the given wavelength and $\int_{-\infty}^{\infty}\mathcal{E}_{\rm led}(\lambda)\mathrm{d}\lambda=1$. In addition, LED and PD elements typically have a low pass characteristic, which can be modelled by a first-order Butterworth filter \cite{7222376}. The impulse response can be written as:
\begin{align}
\mathfrak{h}_{\rm lp}(t)=2\pi f_{\rm c}\mathrm{e}^{-2\pi f_{\rm c}t}\mathfrak{u}(t),
\label{eq:cir_lp}
\end{align}
where $f_{\rm c}$ is the cut-off frequency of the LED or the PD. The frequency response of $\mathfrak{h}_{\rm lp}(t)$ can be found as:
\begin{align}
\mathfrak{H}_{\rm lp}(f)=\frac{1}{1+j\frac{f}{f_{\rm c}}}.
\label{eq:fr_lp}
\end{align}
$\mathfrak{h}_{\rm oc}(t)$ in \eqref{eq:analogue_cir} represents the impulse response of the optical channel, which includes the response due to signal propagation via a line-of-sight (LoS) path and the responses due to non-line-of-sight (NLoS) paths caused by reflections. Infrared LEDs typically have  a low optical power of up to a few hundred mW. Even with multiple LED modules, the combined optical power is still limited to a level that fulfils the photobiological safety standard \cite{std:bsen62471:2008}. In this condition, a reliable uplink transmission will primarily rely on a robust LoS channel. Otherwise, the strength of the received signal is insufficient to overcome the distortion due to receiver noise. Studies show that reflected signals have little effect on the performance of an optical wireless communication system when the optical channel is dominated by a robust LoS path \cite{cbh1601}. In addition, compared to the low pass characteristics caused by front-end devices such as LEDs and PDs, the frequency selectivity caused by reflected signals is negligible. Consequently, we omit the optical channel based on NLoS
paths. Therefore, the impulse response of the optical channel is simplified as:
\begin{align}
\mathfrak{h}_{\rm oc}(t)=G\delta\left(t-\frac{D}{c}\right),
\label{eq:cir_oc}
\end{align}
where $c$ is the speed of light, $D$ is the Euclidean distance between the UE transmitter and the AP receiver, and $G$ is the LoS path loss. The characteristics of $G$ depends on the complicated uplink transmission geometry, which will be treated individually in Section~\ref{sec:LoS_characteristics}. $\mathfrak{h}_{{\rm pd},\lambda}(t)$ in \eqref{eq:analogue_cir} is the impulse response of the PD, which can be decomposed as:
\begin{align}
\mathfrak{h}_{{\rm pd},\lambda}(t)=\mathcal{R}_{\rm pd}(\lambda)\mathfrak{h}_{\rm lp,pd}(t),
\label{eq:cir_pd}
\end{align}
where $\mathcal{R}_{\rm pd}(\lambda)$ is the spectral sensitivity of the PD at wavelength $\lambda$, and $\mathfrak{h}_{\rm lp,pd}(t)$ models the low pass characteristics of the PD, which is defined by \eqref{eq:cir_lp}.

The channel blocks between $x[n]$ and $y[n]$ in Fig.~\ref{fig:comb1}~(a) are analogue, continuous-time systems. However, the input $x[n]$ and output $y[n]$ of the channel are discrete-time signals. Therefore, we can use an equivalent discrete-time impulse response $h[n]$ to represent the characteristics of \eqref{eq:analogue_cir}. The discrete-time impulse response can be calculated as $h[n]=\mathfrak{h}(nT_{\rm s})$, where $T_{\rm s}$ is the symbol period. In conjunction with \eqref{eq:led_cir}, \eqref{eq:fr_lp}, \eqref{eq:cir_oc} and \eqref{eq:cir_pd}, the frequency response at the $k^{\rm th}$ frequency after a $K$-point discrete Fourier transform (DFT) can be found as:
\begin{align}
H[k]=\sigma_{\rm o}\mathcal{F}_{\lambda}GH_{\rm lp}[k]
\label{eq:frequency_response}
\end{align}
for $k=0, 1, \cdots, K-1$, where $\mathcal{F}_{\lambda}=\int_{\lambda_{\rm min}}^{\lambda_{\rm max}}\mathcal{E}_{\rm led}(\lambda)\mathcal{R}_{\rm pd}(\lambda)\mathrm{d}\lambda$ and 
\begin{align}
H_{\rm lp}[k]=\left\{\begin{array}{lr} \frac{1}{\left(1+\frac{jkf_{\rm s}}{Kf_{\rm c,led}}\right)\left(1+\frac{jkf_{\rm s}}{Kf_{\rm c,pd}}\right)} &: k<\frac{K}{2} \\
\frac{1}{\left(1+\frac{j(k-K)f_{\rm s}}{Kf_{\rm c,led}}\right)\left(1+\frac{j(k-K)f_{\rm s}}{Kf_{\rm c,pd}}\right)} &: k\geq\frac{K}{2} \end{array}
\right..
\label{eq:H_k}
\end{align}
With the definition of $h[n]$, the received discrete-time signal samples can be written as:
\begin{align}
y[n]=x[n]\otimes h[n]+w[n],
\end{align}
where $w[n]$ is the receiver noise samples. In this study, the receiver noise is modelled as an additive white Gaussian noise (AWGN) with a variance of $\sigma_{w}^2=\frac{N_0}{2}f_{\rm s}$, where $f_{\rm s}=1/T_{\rm s}$ is the double-side modulation bandwidth and $\frac{N_0}{2}$ is the double-side noise power spectral density (PSD). Note that the modulation bandwidth $f_{\rm s}$ corresponds to the used overall receiver bandwidth, which is also equal to the symbol rate of the transmission. In an optical wireless system, the receiver noise PSD is dominated by the thermal noise and shot noise. However, the uplink receiver is deployed on the ceiling facing downwards, which makes the detection of direct sunlight impossible. In addition, transmission rate in the infrared region of the windows can be reduced to further decrease the effects of background shot noise. Due to the low optical power of the infrared LED, the variance of the signal dependent shot noise is negligible compared to thermal noise. Therefore, the receiver noise is dominated by the thermal noise, and the  noise PSD can be calculated as \cite{Islim:18} $\frac{N_0}{2}=\frac{4\mathcal{K}_{\rm b}\mathcal{T}}{\mathcal{R}_{\rm L}}$, where $\mathcal{K}_{\rm b}$ is the Boltzmann's constant, $\mathcal{T}$ is the absolute temperature and $\mathcal{R}_{\rm L}$ is the load resistance.

\subsection{Pulse Amplitude Modulation (PAM) with Single Carrier Frequency Domain Equalisation (SCFDE)}
\label{subsec:PAM_SCFDE}
In this paper, the PAM-SCFDE has been considered \cite{7089163}. Firstly, PAM has a relatively lower peak-to-average power ratio (PAPR) compared to a high PAPR modulated signal, such as orthogonal frequency division multiplexing (OFDM). This is an important characteristic in a wireless infrared uplink system with a low optical power and small dynamic range. Secondly, compared to a conventional time domain equaliser, SCFDE has a significantly lower computational complexity \cite{4607215}.
The block diagram of PAM-SCFDE is shown in Fig.~\ref{fig:comb1}~(b). In PAM-SCFDE, every $K$ bipolar-PAM symbols are grouped as a transmission block. After the PAM modulator, the discrete-time signal can be defined as: $\tilde{x}[n]=\rho_M a_{\mathfrak{m}}$, where $M$ is the constellation size, $a_{\mathfrak{m}}=2\mathfrak{m}+1-M$ for $\mathfrak{m}=0,1,\cdots,M-1$, and $\rho_M=\sqrt{3/\left(M^2-1\right)}$ is a symbol normalisation factor which makes the variance of $\tilde{x}[n]$ equal to unity. Similar to an OFDM transmission, each block is transmitted, equalised and decoded independently. A cyclic prefix (CP) is added by repeating the last $L_{\rm cp}$ symbols in a transmission block, and placing these symbols at the beginning of that block. It can prevent the interference between adjacent transmission blocks. In addition, it creates the circular convolution characteristic for the convenience of the single-tap frequency domain equalisation. On the receiver side, the received symbols are converted to the samples in the frequency domain by a $K$-point fast Fourier transform (FFT). After the frequency domain equalisation, the recovered samples are converted back to the time domain via a $K$-point inverse fast Fourier transform (IFFT). Finally, the equalised PAM symbols are demodulated and decoded back to binary information. Due to the low-pass effects caused by front-end elements, the received SNR at different frequency varies. In conjunction with \eqref{eq:frequency_response}, the SNR of the samples at $k^{\rm th}$ frequency after the FFT operation can be calculated as:
\begin{align}
\gamma_{k}=\frac{\sigma_{x}^2\left|H[k]\right|^2}{\sigma_{w}^2}=\frac{\Delta P_{\rm o}^2\mathcal{F}_{\rm \lambda}^2G^2\left|H_{\rm lp}[k]\right|^2}{2\eta_{\rm c}^2N_0f_{\rm s}}.
\label{eq:snr_on_k_frequency}
\end{align}
Recall that $\sigma_{x}^2$ is the variance of $x[n]$. The SNR of the equalised PAM signal in the time domain after IFFT operation, $\gamma_{\rm pam}$ is related to value of each $\gamma_{k}$. In the case with linear minimum mean square error (LMMSE) equaliser, $\gamma_{\rm pam}$ can be calculated as \cite{7089163}:
\begin{align}
\gamma_{\rm pam}=\left(\frac{1}{K}\sum_{k=0}^{K-1}\frac{1}{1+\gamma_{k}}\right)^{-1}-1.
\label{eq:SNR_pam}
\end{align} 
In the case of biploar $M$-PAM, the crest factor can be calculated as \cite{book:x0601}:
\begin{align}
\eta_{\rm c}=\sqrt{\frac{3(M-1)}{M+1}}.
\label{eq:pam_crest_factor}
\end{align} 
The bit error rate (BER) of bi-polar $M$-PAM with gray code can be calculated as \cite{book:x0601}:
\begin{align}
\mathcal{P}_{\rm ber}(M,f_{\rm s})=\frac{2(M-1)}{M\log_2M}\mathcal{Q}\left(\sqrt{\frac{3\gamma_{\rm pam}}{M^2-1}}\right).
\label{eq:ber_pam}
\end{align}
Note that according to \eqref{eq:SNR_pam}, \eqref{eq:snr_on_k_frequency} and \eqref{eq:pam_crest_factor}, $\gamma_{\rm pam}$ is a function of $M$ and $f_{\rm s}$, as well. \subsubsection{Achievable data rate with fixed modulation bandwidth}
\label{subsubsec:data_rate_fixed}
We first consider the case with a given modulation bandwidth $f_{\rm s}$. In order to increase the achievable data rate, the largest constellation size that results in an acceptable BER should be used. A target BER is defined as $\mathcal{P}_{\rm ber}^{\rm tar}$. We assume that the possible constellation sizes are drawn from the set $\mathcal{M}=\{2,4,\cdots,M_{\rm max}\}$, where $M_{\rm max}$ is the largest used PAM constellation size. Thus, the set of constellations that can achieve a BER no higher than the target BER can be found as:
$\tilde{\mathcal{M}}=\left\{M\in\mathcal{M}:\mathcal{P}_{\rm ber}(M,f_{\rm s})\leq \mathcal{P}_{\rm ber}^{\rm tar}\right\}$. The achievable data rate  can be written as:
\begin{align}
R_{\rm b,fix}^{\rm pam}=\frac{f_{\rm s}K}{K+L_{\rm cp}}\log_2\left(\max\tilde{\mathcal{M}}\right).
\label{eq:data_rate_fixed}
\end{align}
Note that if $\tilde{\mathcal{M}}=\emptyset$, no data can be transmitted, and $R_{\rm b,1}^{\rm pam}=0$. \subsubsection{Achievable data rate with adaptive modulation bandwidth}
\label{subsubsec:data_rate_adaptive}
In the case that $f_{\rm s}$ can be adjusted, the achievable data rate can be further improved. Considering each constellation size $M$, we maximise $f_{\rm s}$ so that the BER $\mathcal{P}_{\rm ber}$ reaches the target BER $\mathcal{P}_{\rm ber}^{\rm tar}$. Thus, the optimal $f_{\rm s}$ respect to a specified $M$ can be calculated as:
\begin{align}
\hat{f}_{{\rm s},M}=\mathcal{P}_{\rm ber}^{-1}\left(M,\mathcal{P}_{\rm ber}^{\rm tar}\right),
\end{align}
where $\mathcal{P}_{\rm ber}^{-1}$ is the inverse function of \eqref{eq:ber_pam} respect to the variable $f_{\rm s}$, which is not tractable but can be numerically evaluated using the bisection method. Then, the maximum achievable data rate using a constellation size of $M$ can be found as:
\begin{align}
\hat{R}_{{\rm b},M}^{\rm pam}=\frac{\hat{f}_{{\rm s},M}K}{K+L_{\rm cp}}\log_2 M.
\end{align}
Finally, we select the constellation size that can achieve the highest data rate to transmit:
\begin{align}
R_{\rm b,ad}^{\rm pam}=\max\left\{\hat{R}_{{\rm b},2}^{\rm pam}, \hat{R}_{{\rm b},4}^{\rm pam},\cdots,\hat{R}_{{\rm b},M_{\rm max}}^{\rm pam}\right\}.
\label{eq:data_rate_adaptive}
\end{align}

Next, we evaluate the achievable data rate under various channel conditions using PAM-SCFDE. It is intuitive to calculate the data rate against SNR at the receiver. However, it is implied in Section~\ref{subsubsec:data_rate_fixed} and \ref{subsubsec:data_rate_adaptive} that with various channel conditions, different constellation size $M$ and modulation bandwidth $f_{\rm s}$ may be used to maximise the achievable data rate. As shown in \eqref{eq:snr_on_k_frequency} and \eqref{eq:SNR_pam}, the receiver side SNR of PAM-SCFDE is a function of $M$ and $f_{\rm s}$. For the convenience of presentation, the SNR at the $k^{\rm th}$ frequency \eqref{eq:snr_on_k_frequency} is rewritten as $\gamma_{k}=\xi\left|H_{\rm lp}[k]\right|^2/\left(\eta_{\rm c}^2f_{\rm s}\right)$, where
\begin{align}
\xi=\frac{\Delta P_{\rm o}^2\mathcal{F}_{\rm \lambda}^2G^2}{2N_0},
\label{eq:channel_factor}
\end{align}
is defined as a channel factor which encapsulates the characteristics of optical channel and front-end elements. 

\begin{figure}
	\centering
	\begin{subfigure}{.59\textwidth}
		\centering
		\includegraphics[width=1\textwidth]{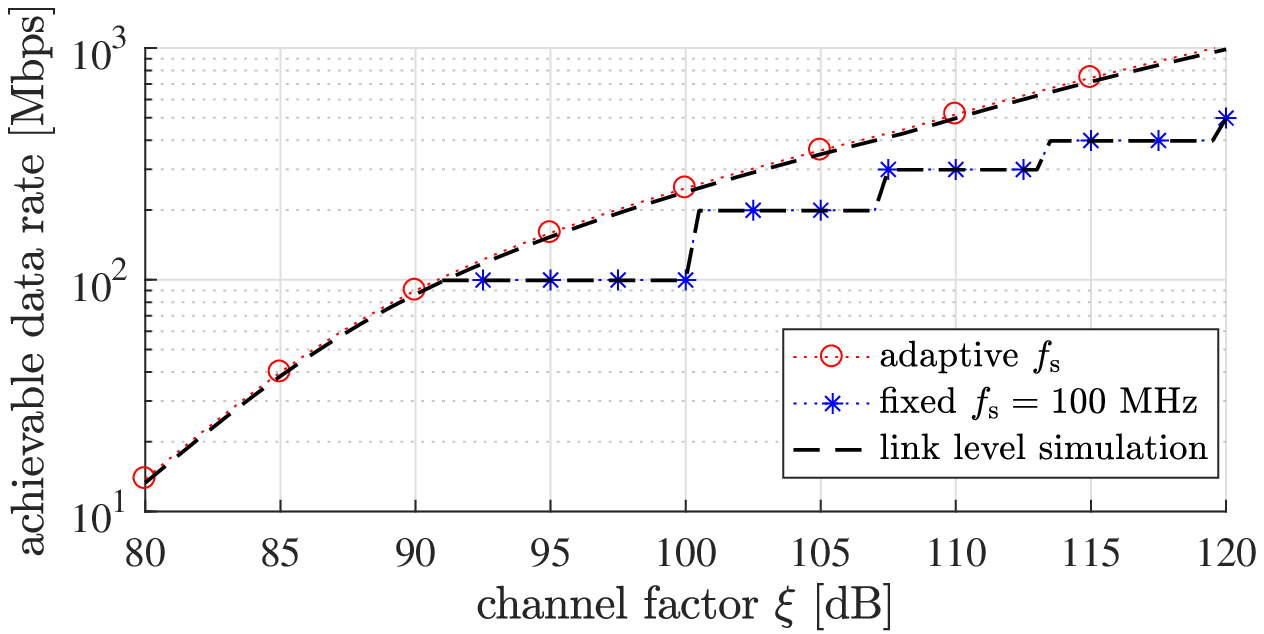}
		\caption{}
		\label{}
	\end{subfigure}	
	\begin{subfigure}{.39\textwidth}
		\centering
		\includegraphics[width=1\linewidth]{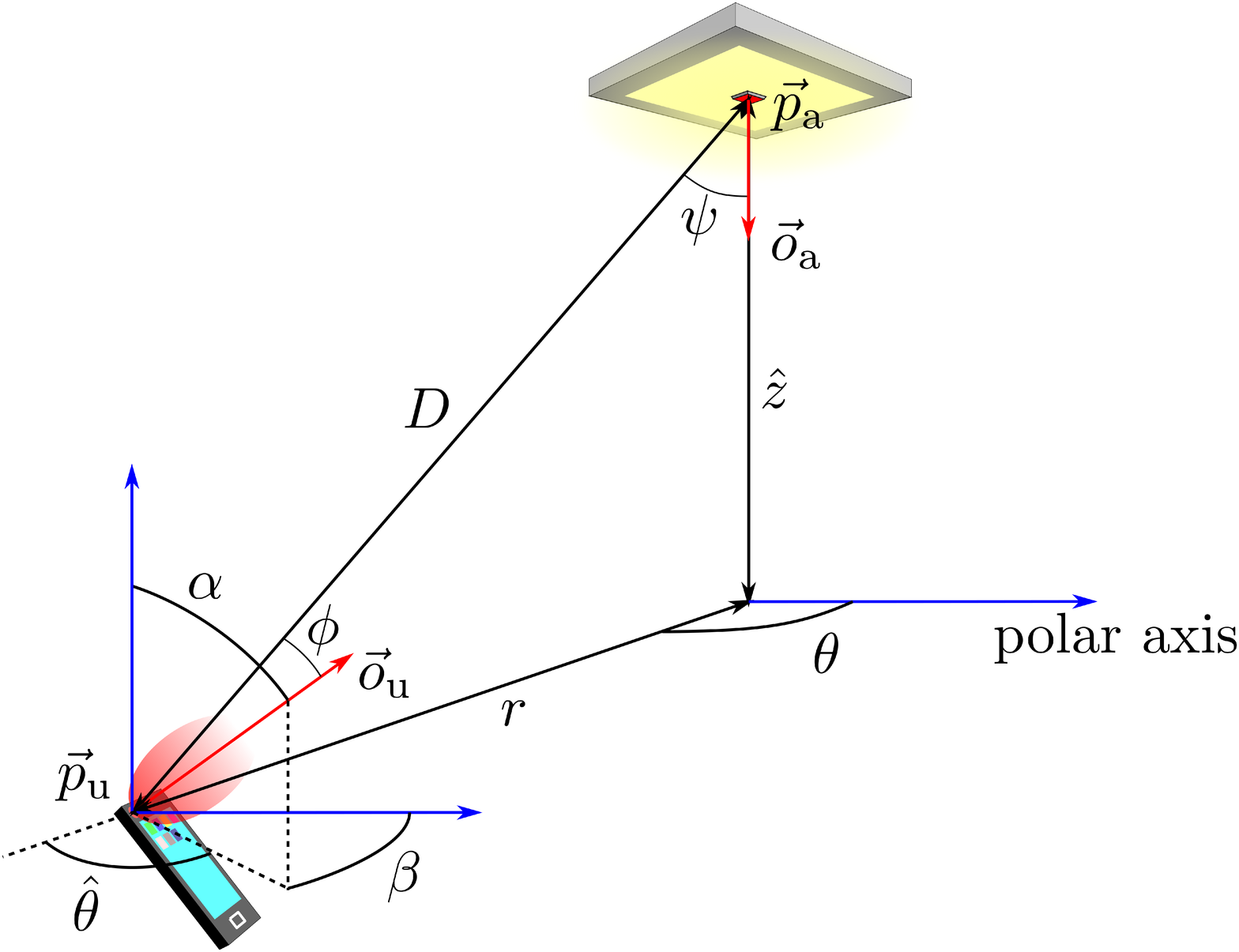}
		\caption{}
		\label{}
	\end{subfigure}%
	\caption{(a) Achievable data rate against channel factor. (b) Wireless infrared uplink optical channel geometry.}
	\label{fig:comb2}
\end{figure}

Based on \eqref{eq:data_rate_fixed} and \eqref{eq:data_rate_adaptive}, the results of the achievable data rates $R_{\rm b,fix}^{\rm pam}$ and $R_{\rm b,ad}^{\rm pam}$ using PAM-SCFDE against the channel factor $\xi$ are calculated and shown in Fig.~\ref{fig:comb2}~(a). Note that the characteristics of \eqref{eq:H_k} depends on the front-end elements. In this study, the infrared LED VSMY3940X01 \cite{VSMY3940X01} and the infrared-enhanced avalanche photodiode (APD) S11519-30 \cite{S11519-30} are used. The related parameters and modulation configurations are listed in Table~\ref{table:parameters1}. From Fig.~\ref{fig:comb2}~(a), we can observe that with an increase of channel factor $\xi$, the achievable data rate increases. Compared to the case with fixed modulation bandwidth $f_{\rm s}=100$~MHz, a higher data rate can be achieved with an optimisation of $f_{\rm s}$. In addition, the data rate cannot improve until the channel factor $\xi$ exceeds a certain level in the case of fixed $f_{\rm s}=100$~MHz.

\section{Uplink Optical Channel Model}
\label{sec:LoS_characteristics}
In Section~\ref{subsec:PAM_SCFDE}, we have shown that the achievable data rate depends on the channel factor $\xi$. According to \eqref{eq:channel_factor}, the value of $\xi$ depends on $\Delta P_{\rm o}$, $\mathcal{F}_{\lambda}$, $N_0$ and $G$. The values of $\Delta P_{\rm o}$ and $\mathcal{F}_{\lambda}$ depend on the chosen front-end elements, which can be improved by selecting optical source and detector with higher speed and higher power/sensitivity. The value of $N_0$ depends on the environment and the receiver's electronics. The value of $G$ is dominated by the uplink geometric characteristics, such as orientations of LED/PD and the separation between a UE and an AP. Therefore, $G$ can be treated as a random variable which causes uncertainty to the value of $\xi$. In this section, we present the model of optical channel as the basis for developing the analytical statistics of $G$. In this study, we consider the randomness arising from UE position, link blockage and UE orientation.
\subsection{Uplink Geometry}
\label{subsec:Uplink Geometry}
Firstly, we consider a LiFi AP and a UE defined in a three-dimensional indoor space as shown in Fig.~\ref{fig:comb2}~(b).The positions of the AP and the UE are initially defined by a cylinder coordinate system as $(0,0,z_{\rm a})$ and $(r,\theta,z_{\rm u})$, respectively. The origin of the coordinate system is right below the position of the LiFi AP. Note that the transmitter is installed on the UE while the receiver is on the AP side in an uplink system. For the analytical convenience, we convert the position vector in the cylinder coordinate system to another position vector in a Cartesian coordinate system as $\vec{p}_{\rm a}=(0,0,z_{\rm a})$ and $\vec{p}_{\rm u}=(r\cos\theta,r\sin\theta,z_{\rm u})$. Then we consider the orientations of the transmitter source and the receiver detector. Due to the fact that all UEs are below the APs and randomly placed in the indoor environment, it is intuitive to assume that the orientation of the AP receiver detector is directed to the floor, which can be defined by a unit vector $\vec{o}_{\rm a}=(0,0,-1)$ in the Cartesian coordinate system. On the other hand, the orientation of a UE is highly dependent on the way that the user holds the mobile device. In this study, we assume that the uplink transmitter source is installed in the front of a mobile device, which is similar to the installation of the front camera on a smart phone. The detector orientation can be defined by two angles in a spherical coordinate system\cite{8540452}: elevation/zenith angle $\alpha$ and azimuth angle $\beta$. The corresponding orientation vector in the Cartesian coordinate system can be written as a unit vector $\vec{o}_{\rm u}=(\cos\beta\sin\alpha,\sin\beta\sin\alpha,\cos\alpha)$.

\begin{table}[!t]
	\caption{System parameters: front-end elements and modulation}
	\centering
	\begin{tabular}{c c c|c c c}
		\hline
		\hline
		Parameter & Symbol & Value & Parameter & Symbol & Value  \\	
		\hline					
		LED maximum optical power & $P_{\rm o,max}$ & 440~mW & PD cut-off frequency & $f_{\rm c,pd}$ & 230~MHz  \\
		LED minimum optical power & $P_{\rm o,min}$ & 0~mW  & FFT size & $K$ & 512 \\	
		LED cut-off frequency & $f_{\rm c,led}$ & 35~MHz &	Target BER & $\mathcal{P}_{\rm b}^{\rm T}$ & $3.8\times 10^{-3}$ \\		
		PD physical area & $A_{\rm pd}$ & 7.1~$\rm mm^2$ & PAM maximum constellation size & $M_{\rm max}$ & 32 \\	
		APD gain & - & 10 & ~ & ~ &	~ \\
		\hline
		\hline
	\end{tabular}
	\label{table:parameters1}
\end{table}
\subsection{Uplink Optical Channel Path Loss}
\label{subsec:LoS channel}
The wireless free space path loss via the LoS transmission path can be calculated using the following expression \cite{kb9701}:
\begin{align}
G=\frac{(m+1)A_{\rm pd}\mathcal{F}_{\rm c}}{2\pi D^2}\mathbf{1}_{\mathcal{V}}\cos^m\phi\cos\psi,
\label{eq:G_1}
\end{align}
where $m$ is the Lambertian mode number which is related to the LED half-power semiangle $\phi_{1/2}$ by $m=-1/\log_2(\cos\phi_{1/2})$, $D$ is the Euclidean norm between the transmitter source and the receiver detector, $\phi$ and $\psi$ are the  corresponding radiant / incident angles, $A_{\rm pd}$ denotes the physical area of the PD, $\mathcal{F}_{\rm c}$ denotes an optical concentrator gain and $\mathbf{1}_{\mathcal{V}}$ is a visibility factor. The value of the concentrator gain is calculated by $\mathcal{F}_{\rm c}=\mathfrak{n}^2/\sin^2\psi_{\rm FoV}$, where $\mathfrak{n}$ is the refractive index of the concentrator material and $\psi_{\rm FoV}$ is the field of view (FoV) of the receiver detector. The visibility factor makes the channel gain equal to zero when either $\phi$ exceed $\pi/2$ or $\psi$ is greater than $\psi_{\rm FoV}$. In conjunction with the geometric vectors defined in Section~\ref{subsec:Uplink Geometry}, the value of $D$, $\cos\phi$, and $\cos\psi$ can be calculated as \cite{bkklm9301}:
\begin{align}
D&=||\vec{p}_{\rm u}-\vec{p}_{\rm a}||=\sqrt{r^2+\hat{z}^2}, \label{eq:D} \\
\cos\phi&=\frac{\vec{o}_{\rm u}\left(\vec{p}_{\rm a}-\vec{p}_{\rm u}\right)^\mathrm{T}}{D}=\frac{ \hat{z}\cos\alpha-r\sin\alpha\cos\hat{\theta}}{\sqrt{r^2+\hat{z}^2}}, \label{eq:cosphi} \\
\cos\psi&=\frac{\vec{o}_{\rm a}\left(\vec{p}_{\rm u}-\vec{p}_{\rm a}\right)^\mathrm{T}}{D}=\frac{\hat{z}}{\sqrt{r^2+\hat{z}^2}}, \label{eq:cospsi}
\end{align}
where $\hat{z}=z_{\rm a}-z_{\rm u}$ is the transmitter-receiver height difference and $\hat{\theta}=\theta-\beta$ is an angle difference between the polar angle of UE position and the UE orientation azimuth angle, as shown in Fig.~\ref{fig:comb2}~(b). By inserting \eqref{eq:D}, \eqref{eq:cosphi} and \eqref{eq:cospsi} into \eqref{eq:G_1}, the channel path loss can be rewritten as:
\begin{align}
G=c_0\mathbf{1}_{\mathcal{V}}\left(\hat{z}\cos\alpha-r\sin\alpha\cos\hat{\theta}\right)^m\left(r^2+\hat{ z}^2\right)^{-\frac{m+3}{2}}, \label{eq:G_2}
\end{align}
where $c_0=\frac{1}{2\pi}(m+1)A_{\rm pd}\mathcal{F}_{\rm c}\hat{z}$ is a constant depends on a few deterministic parameters. By inspecting \eqref{eq:cosphi}, we find that $\cos\phi<0$ if $\hat{z}\cos\alpha-r\sin\alpha\cos\hat{\theta}>0$. By comparing \eqref{eq:cospsi} with $\cos\psi_{\rm FoV}$, the UE is out of the coverage of the AP if $r>\hat{z}\tan\psi_{\rm FoV}$. Thus, the visibility factor can be calculated by an indicator function which is defined as:
\begin{align}
\mathbf{1}_{\mathcal{S}}(x)&=\left\{
\begin{array}{lr}
1&: x\in\mathcal{S} \\ 0&: x\notin\mathcal{S}
\end{array}\right.,
\end{align}
where the set $\mathcal{S}$ corresponds to the set $\mathcal{V}$ in the visibility factor $\mathbf{1}_{\mathcal{V}}$, which can be concluded as: 
\begin{align}
\mathcal{V}=\{r:r\tan\alpha\cos\hat{\theta}<\hat{z} ~\text{or}~r>\hat{z}\tan\psi_{\rm FoV}\}.
\end{align}

\subsection{Blockage by Human Body}
\label{subsec:blockage}
In addition to the visibility factor, link blockages may also lead to the disruption of an optical channel. The people in the indoor environment are the major cause of the blockage. In this study, we model the human blocker as a cylinder object with a height of $z_{\rm b}$ and a radius of $l_{\rm b}$ \cite{dlz1201}, as shown in Fig.~\ref{fig:blockage}. Once the LoS segment has to intersect with the blocking object, the link is blocked and the AP receiver can no longer detect any uplink signal. There are two basic rules for determining whether a blocker is obstructing a specific LoS path with a given transmitter and receiver positions \cite{dlz1201}: 
\begin{enumerate}[i)]
	\item From the side view, the block is high enough to have the possibility to contact the LoS segment. \label{list:block_rule1} 
	\item From the top view, the LoS segment intersects with the blocker circular foot-print on the floor. \label{list:block_rule2} 
\end{enumerate}
\subsubsection{Non-user blocker (NUB)}
\label{subsubsec:nonuser_blockage}
\begin{figure}[t!]
	\centering
	\includegraphics[width=0.4\linewidth]{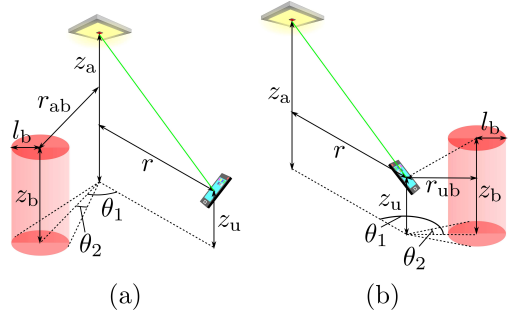}
	\caption{(a) Non-user blocker geometry. (b) user blocker geometry.}
	\label{fig:blockage}
\end{figure}

Two type of blockers are considered in this study. The first type is defined as NUB, which considers the people with a random location who are irrelevant to the considered UEs. As shown in Fig.~\ref{fig:blockage}~(a), the NUB is $r_{\rm ab}$ away from the AP in the horizontal direction. If $r_{\rm ab}\in\left[r\frac{z_{\rm a}-z_{\rm b}}{z_{\rm a}-z_{\rm u}},r\right]$, the blocking rule \ref{list:block_rule1}) is fulfilled \cite{dlz1201}. If $|\theta_1|\leq\theta_2$ or equivalently $|\theta_1|\leq\arcsin(l_{\rm b}/r_{\rm ab})$, the blocking rule \ref{list:block_rule2}) is fulfilled. Assuming that there are $N_{\rm b}$ NUBs randomly distributed in a circular area centred at the AP with a radius of $\mathfrak{r}_{\rm ab}$, the average probability of blockage for a UE that is $r$ away from the AP is concluded as \cite{dlz1201}:
\begin{align}
\hat{\mathcal{P}}_{\rm b}(N_{\rm b},\mathfrak{r}_{\rm ab})=1-\left(1-\frac{2l_{\rm b}r\left(z_{\rm b}-z_{\rm u}\right)}{\pi \mathfrak{r}_{\rm ab}^2\left(z_{\rm a}-z_{\rm u}\right)}\right)^{N_{\rm b}}.
\end{align}
In this study, we assume that the two-dimensional spatial position of the NUBs follows a homogeneous Poisson point process (PPP) with a density of $\Lambda_{\rm b}$. Thus, $N_{\rm b}$ follows a Poisson distribution with a mean of $\Lambda_{\rm b}\pi r^2$. Therefore, the average probability of blockage for a UE that is $r$ away from the AP can be derived as:
\begin{align}
\bar{\mathcal{P}}_{\rm b}&=\sum\limits_{N_{\rm b}=0}^{\infty}\hat{\mathcal{P}}_{\rm b}(N_{\rm b},r)\frac{\left(\Lambda_{\rm b}\pi r^2\right)^{N_{\rm b}}\mathrm{e}^{-\Lambda_{\rm b}\pi r^2}}{N_{\rm b}!}=1-\exp\left(-2l_{\rm b}r\Lambda_{\rm b}\frac{z_{\rm b}-z_{\rm u}}{z_{\rm a}-z_{\rm u}}\right).
\label{eq:NUB_P_b}
\end{align}
\subsubsection{User blocker (UB)}
\label{subsubsec:user blocker}
The second type of blockage considers the case that the signal from a UE is obstructed by its own user. The user of a mobile device is defined as a UB. Because there is a correlation between the positions of the UE and the corresponding UB, its effect should be taken into consideration in the analysis. As shown in Fig.~\ref{fig:blockage}~(b), a UB is $r_{\rm ub}$ away from his UE. If $r\geq\frac{z_{\rm a}-z_{\rm u}}{z_{\rm b}-z_{\rm u}}r_{\rm ub}$, the blocking rule \ref{list:block_rule1}) is fulfilled \cite{fkgm1701}. If $|\theta_1|\leq\theta_2$ or equivalently $|\theta_1|\leq\arcsin(l_{\rm b}/r_{\rm ub})$, the blocking rule \ref{list:block_rule2}) is fulfilled. Note that the second condition is different from that presented in \cite{fkgm1701}, which is due to the fact that a rectangular plate is used as a blocker model in \cite{fkgm1701} while we are using a cylinder as the blocker model. Nevertheless, the blockage characteristics with the two models are similar. We will consider the probability of blockage caused by UBs and the distribution of a LoS channel path loss jointly in the analysis in Section~\ref{sec:channel_path_loss_statistics}. This is because there is a correlation between the UB relative position to the UE and the azimuth angle $\beta$ of the UE orientation.

\section{Optical Channel Path Loss Statistics}
\label{sec:channel_path_loss_statistics}


Based on the model introduced in Section~\ref{sec:LoS_characteristics}, the statistics of the optical channel path loss $G$ are analysed. The considered randomness of $G$ results from the random orientation of the UE and the random probability of blockage. By inspecting \eqref{eq:G_2}, we note that the random variables related to the transmitter orientation include $\alpha$ and $\hat{\theta}$. Regarding the statistics of $\alpha$, we use the experimental results presented in \cite{8540452}, which approximate the statistics of $\alpha$ by a Laplace distribution. The corresponding probability density function (PDF) can be concluded as:
\begin{align}
\mathfrak{f}_{\alpha}(\alpha)=\frac{1}{2b_{\rm L}}\exp\left(-\frac{|\alpha-\mu_{\rm L}|}{b_{\rm L}}\right), 
\label{eq:pdf_alpha}
\end{align}
where $\alpha\in[0,\pi/2]$ and $b_{\rm L}=\sqrt{\sigma_{\rm L}^2/2}$. The value of $\mu_{\rm L}$ and $\sigma_{\rm L}$ was estimated in the experiment carried out in \cite{8540452}. In the case of sitting users, $\mu_{\rm L}=41.06\degree$ and $\sigma_{\rm L}=7.30\degree$. In the case of standing users, $\mu_{\rm L}=29.78\degree$ and $\sigma_{\rm L}=7.87\degree$. The angle $\hat{\theta}$ is the difference between the polar angle $\theta$ and the orientation azimuth angle $\beta$. We assume that there is no preference for these two angles, and both of them have an uniform distribution. Based on this assumption, it can be found that $\hat{\theta}$ also follows a uniform distribution. The corresponding PDF can be written as:
\begin{align}
\mathfrak{f}_{\hat{\theta}}\left(\hat{\theta}\right)=\frac{1}{2\pi},
\label{eq:pdf_hat_theta}
\end{align}
for $\hat{\theta}\in[-\pi,\pi)$. Before determining the distribution of $G$, we consider a few mathematical approximations:
\begin{enumerate}[a)]
	\item Now we consider the inverse trigonometric function $\arccos(u)$. With the function value at $u_0$ known, the function value at $u=u_0+\Delta u$ can be approximated by the first two terms of the function Taylor expansion about $u_0$ as $\arccos(u)\approx\arccos(u_0)-\frac{u-u_0}{\sqrt{1-u_0^2}}$ as long as $\Delta u$ is sufficiently small.
	\item Considering an angle $\omega$, if $\omega\approx0$, the following approximations holds: $\sin \omega \approx \omega$ and $\cos \omega \approx 1$.
\end{enumerate}
%


\newtheorem{theorem}{Theorem}
\begin{theorem}
	\label{theorem:1}
	In the case that $\alpha$ and $\hat{\theta}$ follow the distributions defined by \eqref{eq:pdf_alpha} and \eqref{eq:pdf_hat_theta}, the cumulative distribution function (CDF) of the optical channel path loss $G$ is given by 
	\begin{align}
	&F_G(T)=\left\{\left(\frac{\arccos(\mathcal{Y}_0)}{2\pi}+\frac{\mathcal{Y}_0+\frac{\hat{z}}{r}\tan\mu_{\rm L}}{2\pi\sqrt{1-\mathcal{Y}_0^2}}\right)\mathcal{A}_1-\frac{\frac{\hat{z}}{\cos\mu_{\rm L}}-\left(\frac{T}{c_0}\right)^{\frac{1}{m}}\left(r^2+\hat{z}^2\right)^{\frac{3}{2m}+\frac{1}{2}}}{2 \pi b_{\rm L}r\cos\mu_{\rm L}\sqrt{1-\mathcal{Y}_0^2}}\mathcal{A}_2\right. \nonumber \\ &+\frac{\mathbf{1}_{\mathcal{S}_1}}{\pi}\arcsin\left(\frac{l_{\rm b}}{r_{\rm ub}}\right)+  \left[\frac{\mathbf{1}_{\mathcal{S}_2}+1}{2}+\frac{\mathbf{1}_{\mathcal{S}_2}\mathrm{sgn}(\tilde{\alpha}_{\rm min})}{2}\left(1-\mathrm{e}^{-\frac{|\tilde{\alpha}_{\rm min}|}{b_{\rm L}}}\right)-\frac{\mathrm{sgn}(\tilde{\alpha}_{\rm max})}{2}\left(1-\mathrm{e}^{-\frac{|\tilde{\alpha}_{\rm max}|}{b_{\rm L}}}\right)\right] \nonumber\\ 
	&\left.\times\left[1-\frac{\mathbf{1}_{\mathcal{S}_1}}{\pi}\arcsin\left(\frac{l_{\rm b}}{r_{\rm ub}}\right)\right] \right\}\mathrm{e}^{{-2l_{\rm b}r\Lambda_{\rm b}\frac{z_{\rm b}-z_{\rm u}}{\hat{z}}}}+1-\mathrm{e}^{{-2l_{\rm b}r\Lambda_{\rm b}\frac{z_{\rm b}-z_{\rm u}}{\hat{z}}}},
	\label{eq:CDF_G_r}
	\end{align}	
	where 
	\begin{align}
	&\tilde{\alpha}_{\rm min}=\left|\arcsin\left[\frac{\left(\frac{T}{c_0}\right)^{\frac{1}{m}}\left(r^2+\hat{z}^2\right)^{\frac{3}{2m}}}{ \left(1-\frac{\mathbf{1}_{\mathcal{S}_1}\mathbf{1}_{\mathcal{S}_2}r^2l_{\rm b}^2}{r_{\rm ub}^2\left(r^2+\hat{z}^2\right)}\right)^{\frac{1}{2}}}\right]-\arctan\left[\frac{\hat{z}}{r}\left(1-\frac{\mathbf{1}_{\mathcal{S}_1}\mathbf{1}_{\mathcal{S}_2}l_{\rm b}^2}{r_{\rm ub}^2}\right)^{-\frac{1}{2}}\right]\right|-\mu_{\rm L}, \label{eq:t_alpha_min} \\
	&\tilde{\alpha}_{\rm max}=\pi-\arcsin\left[\frac{\left(\frac{T}{c_0}\right)^{\frac{1}{m}}\left(r^2+\hat{z}^2\right)^{\frac{3}{2m}}}{ \left(1-\frac{\mathbf{1}_{\mathcal{S}_1}r^2l_{\rm b}^2}{r_{\rm ub}^2\left(r^2+\hat{z}^2\right)}\right)^{\frac{1}{2}}}\right]-\arctan\left[\frac{\hat{z}}{r}\left(1-\frac{\mathbf{1}_{\mathcal{S}_1}l_{\rm b}^2}{r_{\rm ub}^2}\right)^{-\frac{1}{2}}\right]-\mu_{\rm L}, \label{eq:t_alpha_max} \\
	& \mathcal{Y}_0=\left\{\begin{array}{lr} -0.5 &: 0\leq T\leq \mathcal{G}_1 \\ \frac{\left(\zeta(r)+0.5\right)(T-\mathcal{G}_1)}{\mathcal{G}_2-\mathcal{G}_1}-0.5 &: \mathcal{G}_1<T<\mathcal{G}_{\rm th} \\ 0 &: T\geq \mathcal{G}_{\rm th}~\text{or}~T<0 \end{array} \right., \\
	&\mathcal{A}_1=\mathrm{e}^{\frac{[\tilde{\alpha}_{\rm max}]^{-}}{b_{\rm L}}}-\mathrm{e}^{\frac{[\tilde{\alpha}_{\rm min}]^{-}}{b_{\rm L}}}-\mathrm{e}^{-\frac{[\tilde{\alpha}_{\rm max}]^{+}}{b_{\rm L}}}+\mathrm{e}^{-\frac{[\tilde{\alpha}_{\rm min}]^{+}}{b_{\rm L}}}, \label{eq:A1} \\
	&\mathcal{A}_2=\mathrm{e}^{\frac{\tan\mu_{\rm L}}{b_{\rm L}}}\left\{\mathrm{Ei}\left[-\frac{[\tilde{\alpha}_{\rm max}]^{+}+\tan\mu_{\rm L}}{b_{\rm L}}\right]-\mathrm{Ei}\left[-\frac{[\tilde{\alpha}_{\rm min}]^{+}+\tan\mu_{\rm L}}{b_{\rm L}}\right]\right\} \nonumber \\ &+\mathrm{e}^{-\frac{\tan\mu_{\rm L}}{b_{\rm L}}}\left\{\mathrm{Ei}\left[\frac{[\tilde{\alpha}_{\rm max}]^{-}+\tan\mu_{\rm L}}{b_{\rm L}}\right]-\mathrm{Ei}\left[\frac{[\tilde{\alpha}_{\rm min}]^{-}+\tan\mu_{\rm L}}{b_{\rm L}}\right]\right\}.
	\label{eq:A2}
	\end{align}

	In \eqref{eq:CDF_G_r}, \eqref{eq:t_alpha_min} and \eqref{eq:t_alpha_max}, it is defined that $\arcsin(x)=\pi/2$ if $x\geq 1$ and $\arcsin(x)=-\pi/2$ if $x\leq -1$. The calculation of the required minor expressions are concluded in Table~\ref{table:minor_expressions}.
\end{theorem}
\begin{IEEEproof}
	Please refer to Appendix~\ref{app:proof of theorem 1}.
\end{IEEEproof}	
The corresponding PDF can be obtained by calculating the derivative of \eqref{eq:CDF_G_r} as:
\begin{align}
&\mathfrak{f}_{G}(T)=\left\{\frac{\mathrm{e}^{-\frac{|\tilde{\alpha}_{\rm max}|}{b_{\rm L}}}}{2\pi b_{\rm L}}\left(\frac{\hat{z}\left(\tilde{\alpha}_{\rm max}\sin\mu_{\rm L}-\cos\mu_{\rm L}\right)+\left(T/c_0\right)^{\frac{1}{m}}{\left(r^2+\hat{z}^2\right)^{\frac{3}{2m}+\frac{1}{2}}}}{r\left(\tilde{\alpha}_{\rm max}\cos\mu_{\rm L}+\sin\mu_{\rm L}\right)\sqrt{1-\mathcal{Y}_0^2}}+\mathbf{1}_{\mathcal{S}_1}\arcsin\left(\frac{l_{\rm b}}{r_{\rm ub}}\right)\right.\right. \nonumber \\ &\left.+\frac{\mathcal{Y}_0}{\sqrt{1-\mathcal{Y}_0^2}}-\pi+\arccos(\mathcal{Y}_0)\right)\frac{\mathrm{d}\tilde{\alpha}_{\rm max}}{\mathrm{d}T}-\frac{\mathrm{e}^{-\frac{|\tilde{\alpha}_{\rm min}|}{b_{\rm L}}}}{2\pi b_{\rm L}}\left(\frac{\mathcal{Y}_0}{\sqrt{1-\mathcal{Y}_0^2}}-\mathbf{1}_{\mathcal{S}_2}\left[\pi-\mathbf{1}_{\mathcal{S}_1}\arcsin\left(\frac{l_{\rm b}}{r_{\rm ub}}\right)\right] \right.\nonumber \\ &\left.+\frac{\hat{z}\left(\tilde{\alpha}_{\rm min}\sin\mu_{\rm L}-\cos\mu_{\rm L}\right)+\left(T/c_0\right)^{\frac{1}{m}}{\left(r^2+\hat{z}^2\right)^{\frac{3}{2m}+\frac{1}{2}}}}{r\left(\tilde{\alpha}_{\rm min}\cos\mu_{\rm L}+\sin\mu_{\rm L}\right)\sqrt{1-\mathcal{Y}_0^2}}+\arccos(\mathcal{Y}_0)\right)\frac{\mathrm{d}\tilde{\alpha}_{\rm min}}{\mathrm{d}T}+\frac{\mathcal{Y}_0}{2\pi\left(1-\mathcal{Y}_0^2\right)^{\frac{3}{2}}} \nonumber \\ &\times\left(\left(\mathcal{Y}_0+\frac{\hat{z}}{r}\tan\mu_{\rm L}\right)\mathcal{A}_1-\frac{\hat{z}-\left(T/c_0\right)^{\frac{1}{m}}{\left(r^2+\hat{z}^2\right)^{\frac{3}{2m}+\frac{1}{2}}\cos\mu_{\rm L}}}{b_{\rm L}r\cos^2\mu_{\rm L}}\mathcal{A}_2\right)\frac{\mathrm{d}\mathcal{Y}_0}{\mathrm{d}T} \nonumber \\ &\left.+\frac{\left(T/c_0\right)^{\frac{1}{m}}\left(r^2+\hat{z}^2\right)^{\frac{3}{2m}+\frac{1}{2}}\mathcal{A}_2}{2\pi b_{\rm L}rmT\sqrt{1-\mathcal{Y}_0^2}\cos\mu_{\rm L}}\right\}\exp\left({-2l_{\rm b}r\Lambda_{\rm b}\frac{z_{\rm b}-z_{\rm u}}{\hat{z}}}\right)+F_G(0)\delta(T), \label{eq:PDF_G_r}
\end{align}
\begin{table}[!t]
	\caption{Calculation of Minor Expressions.}
	\centering
	\begin{tabular}{l|l}
		\hline
		\hline
		$\mathcal{S}_1=\left\{r:r\geq\frac{\hat{z}}{z_{\rm b}-z_{\rm u}}r_{\rm ub}\right\}$ & $\mathcal{G}_1=c_0\left(\hat{z}\cos\mu_{\rm L}-r\sin\mu_{\rm L}\right)^m\left(r^2+\hat{z}^2\right)^{-\frac{m+3}{2}}$ \\ 
		$\mathcal{S}_2=\left\{T:T>\mathcal{G}_{0}\right\}$ & $\mathcal{G}_2=c_0\left(\hat{z}\cos\mu_{\rm L}+r\sin\mu_{\rm L}\right)^m\left(r^2+\hat{z}^2\right)^{-\frac{m+3}{2}}$ \\
		$\zeta(r)=0.04942\mathrm{e}^{-0.2393r}-0.8925\mathrm{e}^{-5.159r}+0.5718$ &	$\mathcal{G}_{\rm th}=1.45\left(\mathcal{G}_2-\mathcal{G}_1\right)/\left(\zeta(r)+0.5\right)+\mathcal{G}_1$ \\
		$\mathcal{G}_0=c_0\hat{z}^m\left(r^2+\hat{z}^2\right)^{-\frac{m+3}{2}}$ & \\
		\hline
		\hline
	\end{tabular}
	\label{table:minor_expressions}
\end{table}
where
\begin{align}
\frac{\mathrm{d}\tilde{\alpha}_{\rm min}}{\mathrm{d}T}&=\frac{T^{\frac{1}{m}-1}}{m c_0^{\frac{1}{m}}}\left(r^2+ \hat{z}^2\right)^{\frac{3}{2m}}\mathrm{sgn}\left(\mathcal{G}_0-T\right)\left[1-\frac{T^{\frac{2}{m}}\left(r^2+\hat{z}^2\right)^{\frac{3}{m}}}{c_0^{\frac{2}{m}}}-\frac{\mathbf{1}_{\mathcal{S}_1}\mathbf{1}_{\mathcal{S}_2}r^2l_{\rm b}^2}{\left(r^2+\hat{z}^2\right)r_{\rm ub}^2}\right]^{-\frac{1}{2}}, \\
\frac{\mathrm{d}\tilde{\alpha}_{\rm max}}{\mathrm{d}T}&=-\frac{T^{\frac{1}{m}-1}}{m c_0^{\frac{1}{m}}}\left(r^2+ \hat{z}^2\right)^{\frac{3}{2m}}\left[1-\frac{T^{\frac{2}{m}}\left(r^2+\hat{z}^2\right)^{\frac{3}{m}}}{c_0^{\frac{2}{m}}}-\frac{\mathbf{1}_{\mathcal{S}_1}r^2l_{\rm b}^2}{\left(r^2+\hat{z}^2\right)r_{\rm ub}^2}\right]^{-\frac{1}{2}}, \\
\frac{\mathrm{d}\mathcal{Y}_0}{\mathrm{d}T}&=\left\{\begin{array}{lr} \frac{\left(\zeta(r)+0.5\right)}{\mathcal{G}_2-\mathcal{G}_1} &: \mathcal{G}_1\leq T\leq\mathcal{G}_{\rm th} \\ 0 &: \text{otherwise} \end{array} \right..
\end{align}
Note that \eqref{eq:CDF_G_r} and \eqref{eq:PDF_G_r} hold when the threshold $T$ fulfils the condition $0\leq T \leq\mathcal{G}_{\rm max}$ where $\mathcal{G}_{\rm max}$ is defined as:
\begin{align}
\mathcal{G}_{\rm max}=\left\{\begin{array}{lr}
c_0\left(r^2+\hat{z}^2\right)^{-\frac{3}{2}} &:r\notin\mathcal{S}_1 \\ c_0\frac{\left(\hat{z}\cos\mu_{\rm L}+r\sin\mu_{\rm L}\sqrt{1-l_{\rm b}^2/r_{\rm ub}^2}\right)^{m}}{\left(r^2+\hat{z}^2\right)^{\frac{m+3}{2}}} &:r\in\mathcal{S}_1
\end{array}\right..
\end{align}
Otherwise, $\mathfrak{f}_{G}(T)=0$ and $F_{G}(T)=0$ if $T< 0$; $\mathfrak{f}_{G}(T)=0$ and $F_{G}(T)=1$ if $T> \mathcal{G}_{\rm max}$.

\begin{figure}
	\centering
	\begin{subfigure}{.49\textwidth}
		\centering
		\includegraphics[width=1\textwidth]{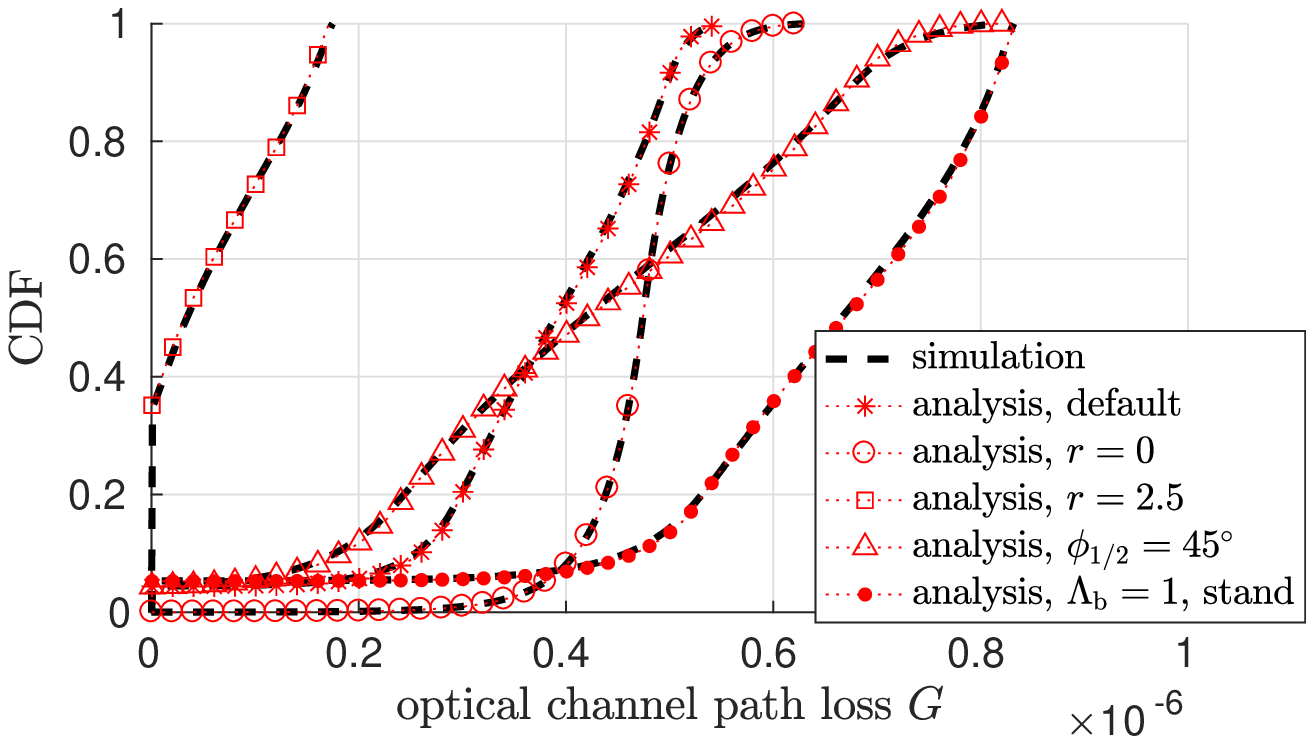}
		\caption{}
		\label{}
	\end{subfigure}	
	\begin{subfigure}{.49\textwidth}
		\centering
		\includegraphics[width=1\linewidth]{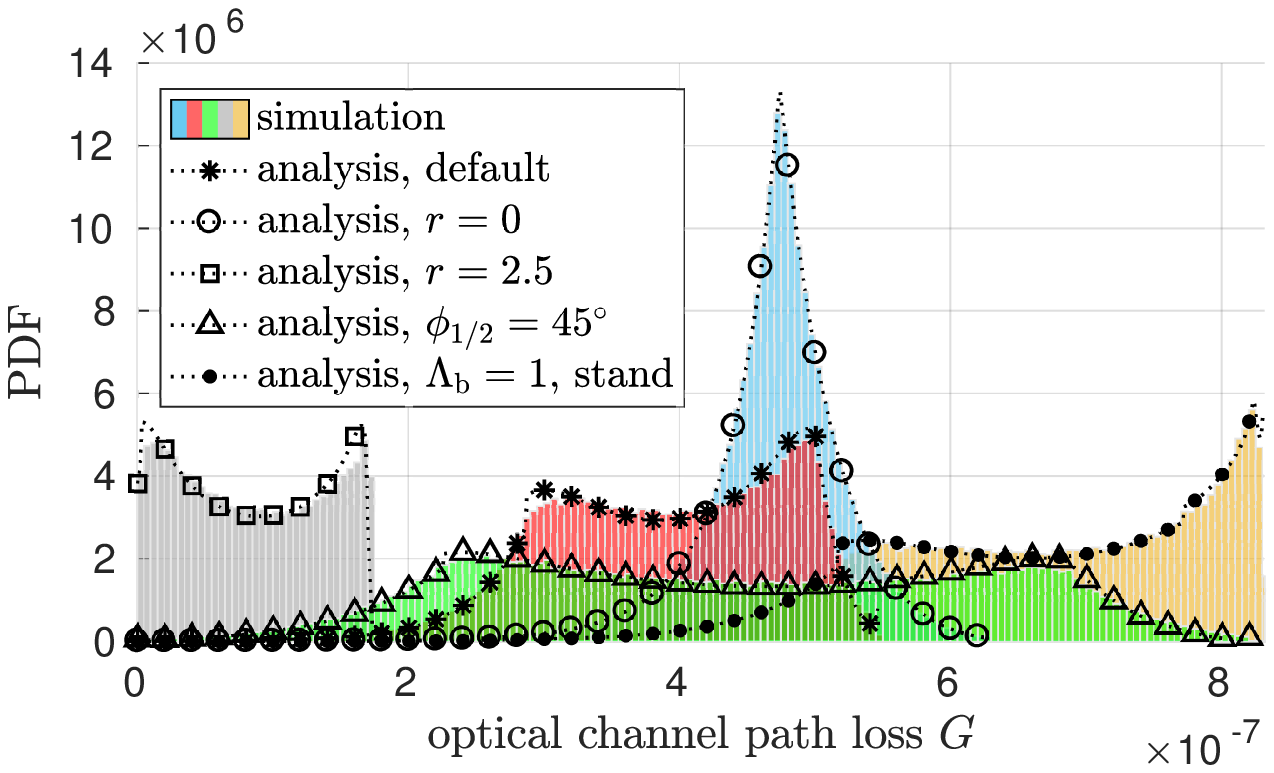}
		\caption{}
		\label{}
	\end{subfigure}%
	\caption{(a) Optical channel path loss CDF. (b) Optical channel path loss PDF.}
	\label{fig:comb3}
\end{figure}

\newtheorem{corollary}{Corollary}
\begin{corollary}
	\label{corollary:2}
	In the case that $\alpha$ and $\hat{\theta}$ follow the distributions defined by \eqref{eq:pdf_alpha} and \eqref{eq:pdf_hat_theta}, the CDF of the optical channel path loss $G$ with a special case of $r=0$ is given by
	\begin{align}
	&F_G(T|r=0)=\frac{1}{2}-\frac{1}{2}\mathrm{sgn}\left[\arccos\left(\left(\frac{T\hat{z}^3}{c_0}\right)^{\frac{1}{m}}\right)-\mu_{\rm L}\right]\left(1-\mathrm{e}^{-\frac{\left|\arccos\left(\left(T\hat{z}^3/c_0\right)^{\frac{1}{m}}\right)-\mu_{\rm L}\right|}{b_{\rm L}}}\right). \label{eq:G_CDF_r0}
	\end{align}
	The corresponding PDF is given by
	\begin{align}
	\mathfrak{f}_G(T|r=0)=\frac{1}{2b_{\rm L}mT}\left({\left(\frac{c_0}{T\hat{z}^3}\right)^{\frac{2}{m}}-1}\right)^{-\frac{1}{2}}\exp\left({-\frac{\left|\arccos\left(\left(T\hat{z}^3/c_0\right)^{\frac{1}{m}}\right)-\mu_{\rm L}\right|}{b_{\rm L}}}\right). \label{eq:G_PDF_r0}
	\end{align}
\end{corollary}
\begin{IEEEproof}
	In the case of $r=0$, the optical channel path loss \eqref{eq:G_2} becomes $G=c_0/\hat{z}^3\cos^m\alpha$. Then we have:
	\begin{align}
	\mathbb{P}[G\leq T]=1-\mathbb{P}\left[\alpha<\arccos\left(\left(\frac{T\hat{z}^3}{c_0}\right)^{\frac{1}{m}}\right)\right].
	\end{align}
	By using the CDF expression of Laplace distribution, the result in \eqref{eq:G_CDF_r0} is obtained. Calculating the derivative of \eqref{eq:G_CDF_r0} leads to the PDF result in \eqref{eq:G_PDF_r0}.
\end{IEEEproof}	
The statistical results calculated using \eqref{eq:G_CDF_r0} and \eqref{eq:G_PDF_r0} are demonstrated in Fig.~\ref{fig:comb3}~(a) and Fig.~\ref{fig:comb3}~(b). The default configuration is given as: $r=0.7$~m, $\phi_{1/2}=60\degree$, $\Lambda_{\rm b}=0.5$ and sitting user is considered. The agreement between the analytical results and the simulation results with various configurations validates the analysis. According to the observation in \cite{8540452}, the distribution of $G$ with randomness of $\alpha$ only follows a distribution similar to a Laplace distribution. From Fig.~\ref{fig:comb4}~(b), we can observe that with the additional randomness of $\theta$, the PDF of $G$ is `broadened' and the distribution has an increased variance. It is also found that $F_{G}(0)$ has non-zero value in some cases. This is because the optical channel has a probability to be obstructed by human blockers in these cases. This blockage probability causes an impulse in the PDF of $G$ at $T=0$, as shown in the last term of \eqref{eq:PDF_G_r}. However, it is not shown in Fig.~\ref{fig:comb4}~(b) to avoid confusion. 

\section{Path loss Statistics and Average Achievable Data Rate}
\label{sec:SNR_data_rate}
In this section, we are going to use the optical channel path loss statistics to find the distribution of the channel factor and the corresponding average achievable data rate. In a multicell network, the layout of APs is crucial to the performance of both downlink and uplink transmissions. With a grid-lattice AP layout, such as square grid deployment, optimistic performance results can be obtained \cite{abg1001}. With a random spatial AP layout, such as PPP deployment, pessimistic performance results can be obtained. The same rule also applies to the LiFi networks \cite{cbh1601}. In this study, we assume that the layout of the APs  follows a homogeneous PPP, since the analysis to the network based on this AP layout has better tractability. For each UE, there are multiple nearby APs. It is intuitive to let the UE to associate with the AP that offers the highest $G$. The path loss corresponds to the associated AP is denoted as $G_0$.


\subsection{Channel factor statistics}
As discussed in Section~\ref{subsec:PAM_SCFDE}, different modulation bandwidths and constellation sizes are used in various channel conditions in order to maximise the achievable data rate. Consequently, the expression for SNR calculation varies with channel condition. Therefore, we defined another parameter, named as channel factor $\xi$ which is independent of the modulation bandwidth and constellation size. We have shown that with a higher channel factor, the achievable data rate increases. Therefore, it is meaningful to determine the statistics of a channel factor. With a given AP receiver FoV $\psi_{\rm FoV}$, the UE can never deliver the signal to an AP, if the horizontal separation between the UE and that AP is greater than $r_{\rm max}=\Delta z\tan\psi_{\rm FoV}$, as the incident angle to the AP exceeds the receiver FoV $\psi_{\rm FoV}$. In the previous sections, $r$ is treated as a deterministic parameter. Taking the random user location into consideration, $r$ is considered to be a random variable in this section. From the point of view of a UE, the candidates for the AP that is connected to the UE are randomly distributed in a circle that has a radius of $r_{\rm max}$ and is centred at the UE. Therefore, the PDF of $r$ can be calculated as $\mathfrak{f}_r(r)=\frac{2r}{r_{\rm max}^2}$, where  $0 \leq r \leq r_{\rm max}$ In conjunction with \eqref{eq:G_CDF_r0}, the CDF of $G$ with respect to one random AP within the circular area can be calculated as $F_G(T)=\int_{0}^{r_{\rm max}}\mathfrak{f}_r(r)F_G(T|r)\mathrm{d}r$. If there are $N_{\rm a}$ APs in the circular area, the CDF of the maximum $G$ can be calculated as $\left\{F_G(T)\right\}^{N_{\rm a}}$ \cite{book:hh03}. Assuming an AP density of $\Lambda_{\rm a}$, the Possion distribution of the number of APs within the circular area can be written as: $\mathbb{P}(N_{\rm a}=l)=\mathrm{e}^{-\Lambda_{\rm a}\pi r_{\rm max}^2}\left(\Lambda_{\rm a}\pi r_{\rm max}^2\right)^l/l!$.
Taking the cases with various number of APs existing in the circular area into account, the CDF of the maximum path loss $G_0$ can be calculated as:
\begin{align}
&F_{G_0}(T)=\sum_{l=0}^{\infty}\mathbb{P}(N_{\rm a}=l)\left(\int\limits_{0}^{r_{\rm max}}\mathfrak{f}_r(r)F_G(T|r)\mathrm{d}r\right)^l =\mathrm{e}^{\int_{0}^{r_{\rm max}}2\pi r\Lambda_{\rm a}F_G(T|r)\mathrm{d}r-\Lambda_{\rm a}\pi r_{\rm max}^2}.
\end{align}
The CDF of the channel factor can be evaluated via the CDF of $G_0$:
\begin{align}
&F_{\xi}(T)=\mathbb{P}\left[\xi\leq T\right]=\mathbb{P}\left[G_0\leq \frac{\sqrt{2TN_0}}{\Delta P_{\rm o}\mathcal{F}_{\rm \lambda}}\right]=\mathrm{e}^{\int_{0}^{r_{\rm max}}2\pi r\Lambda_{\rm a}F_G\left(\frac{\sqrt{2TN_0}}{\Delta P_{\rm o}\mathcal{F}_{\rm \lambda}}\big|r\right)\mathrm{d}r-\Lambda_{\rm a}\pi r_{\rm max}^2}.
\label{eq:nor_SNR_cdf}
\end{align}
The corresponding PDF is the derivative of \eqref{eq:nor_SNR_cdf}:
\begin{align}
&\mathfrak{f}_{\xi}(T)=\frac{\int_{0}^{r_{\rm max}}2\pi r\Lambda_{\rm a}\mathfrak{f}_G\left( \frac{\sqrt{2TN_0}}{\Delta P_{\rm o}\mathcal{F}_{\rm \lambda}}\Big|r\right)\mathrm{d}r-\Lambda_{\rm a}\pi r_{\rm max}^2}{(2T/N_0)^{1/2}\Delta P_{\rm o}\mathcal{F}_{\rm \lambda}}\mathrm{e}^{\int_{0}^{r_{\rm max}}2\pi r\Lambda_{\rm a}F_G\left(\frac{\sqrt{2TN_0}}{\Delta P_{\rm o}\mathcal{F}_{\rm \lambda}}\big|r\right)\mathrm{d}r-\Lambda_{\rm a}\pi r_{\rm max}^2}.
\end{align}
\begin{figure}
	\centering
	\begin{subfigure}{.49\textwidth}
		\centering
		\includegraphics[width=1\textwidth]{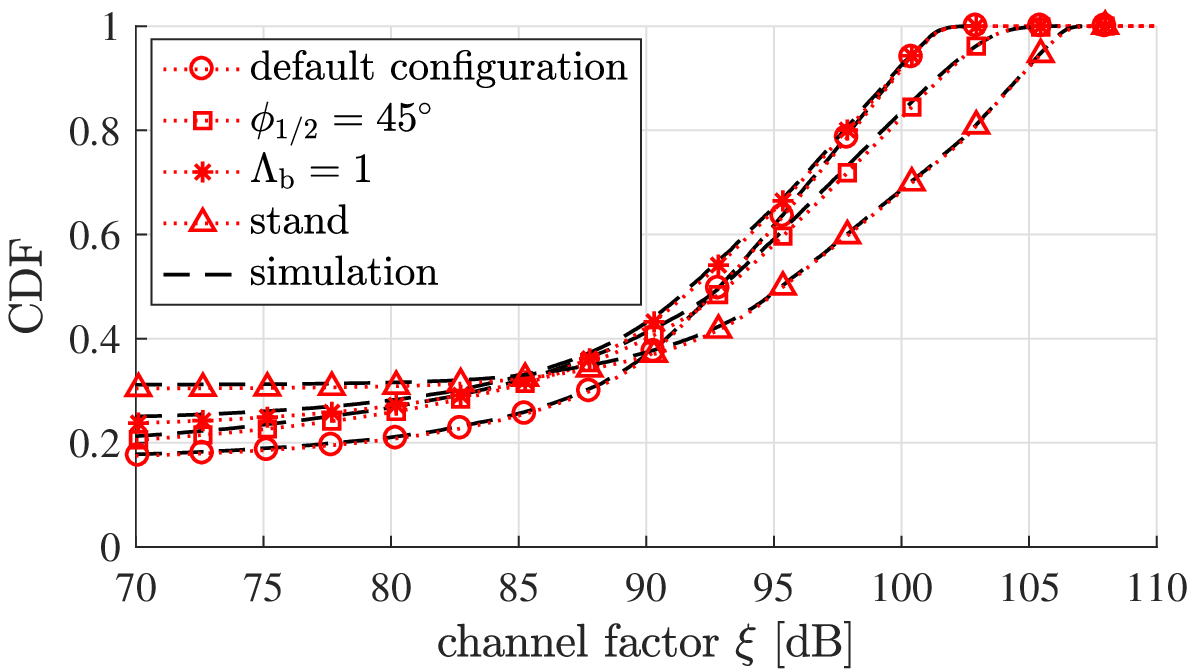}
		\caption{}
		\label{}
	\end{subfigure}	
	\begin{subfigure}{.49\textwidth}
		\centering
		\includegraphics[width=1\linewidth]{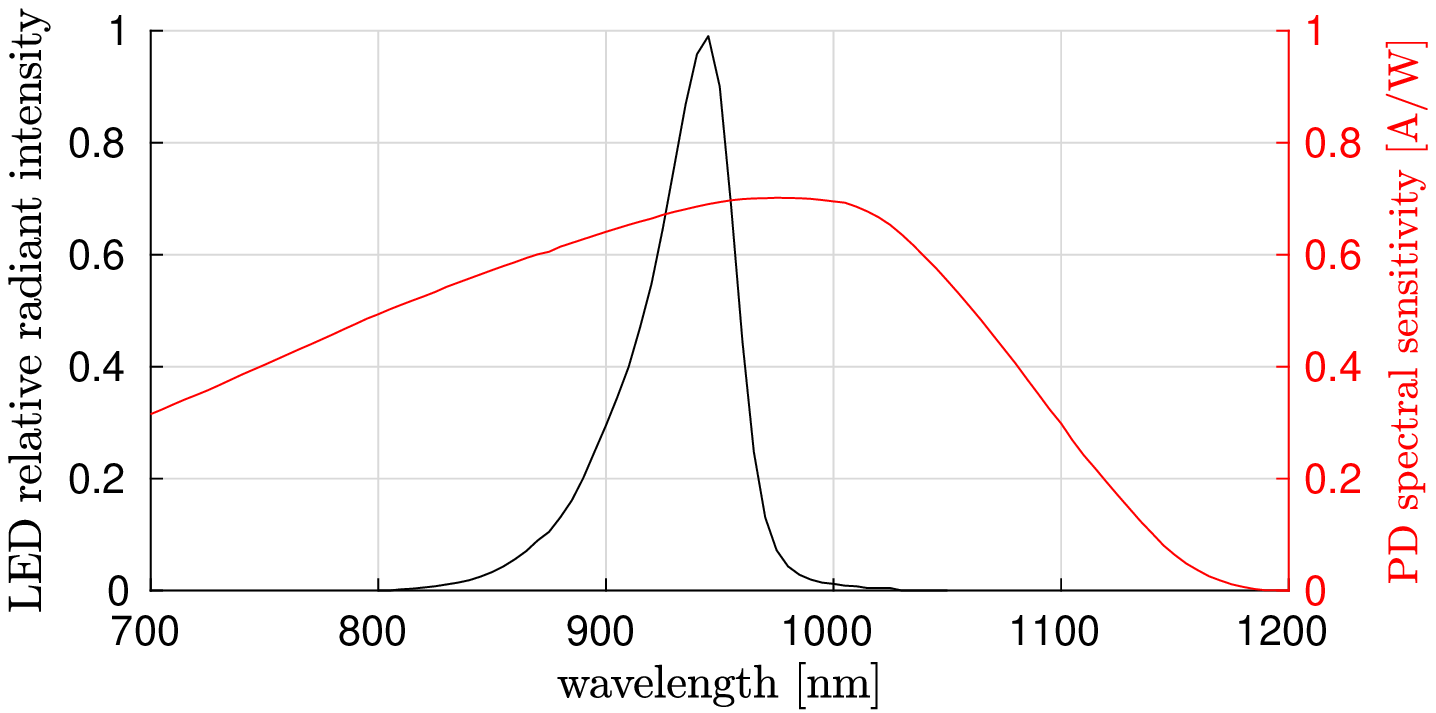}
		\caption{}
		\label{}
	\end{subfigure}%
	\caption{(a) CDF of channel factor $\xi$ with various configurations to $\phi_{1/2}$, $\Lambda_{\rm b}$ and user poses. (b) LED relative radiant intensity and APD spectral sensitivity without APD gain.}
	\label{fig:comb4}
\end{figure}
\begin{table}[!t]
	\caption{System parameters: geometry and noise PSD}
	\centering
	\begin{tabular}{c c c|c c c}
		\hline
		\hline
		Parameter & Symbol & Value & Parameter & Symbol & Value  \\
		\hline
		LED half-power semi-angle & $\phi_{1/2}$ & 60\degree & blocker radius & $l_{\rm b}$ & 0.15~m \\
		receiver FoV & $\psi_{\rm FoV}$ & 50\degree & UE-blocker separation & $r_{\rm ub}$ & 0.3~m \\
		refractive index & $\mathfrak{n}$ & 1.5 & absolute temperature & $\mathcal{T}$ & 290~K \\
		AP height & $z_{\rm a}$ & 3~m & receiver load resistor & $\mathcal{R}_{\rm L}$ & 50~$\Omega$ \\
		UE height (sitting/standing) & $z_{\rm u}$ & 0.75~/~1.25~m & blocker density & $\Lambda_{\rm b}$ & 0.1~blocker/$\rm m^2$ \\
		blocker height & $z_{\rm b}$ & 1.7~m &AP density & $\Lambda_{\rm a}$ & 0.1~AP/${\rm m^2}$  \\
		\hline
		\hline
	\end{tabular}
	\label{table:parameters2}
\end{table}
Fig.~\ref{fig:comb4}~(a) shows the numerical results of \eqref{eq:nor_SNR_cdf} with a number of configurations along with the corresponding Monte Carlo simulation results. If a parameter is not specified, the values listed in Table~\ref{table:parameters2} are used as a default configuration. The spectral dependent parameters related to LED and PD are shown in Fig.~\ref{fig:comb4}~(b). Sitting users are considered in the numerical and simulations results if it is not specified. Fig.~\ref{fig:comb4}~(a) demonstrates that with various configuration of $\phi_{1/2}$, $\Lambda_{\rm b}$ and user status (sitting or standing), the numerical results of the analysis are able to match the corresponding simulation results. The achievable medium normalised SNR is in the range of 90 to 95~dB, and the maximum normalised SNR is in the range of 100 to 105~dB. There is a probability of 20\% to 30\% that the UE can receive no desired signal from any of the nearby APs, which means the user is in outage state.

\begin{figure}
	\centering
	\begin{subfigure}{.49\textwidth}
		\centering
		\includegraphics[width=1\textwidth]{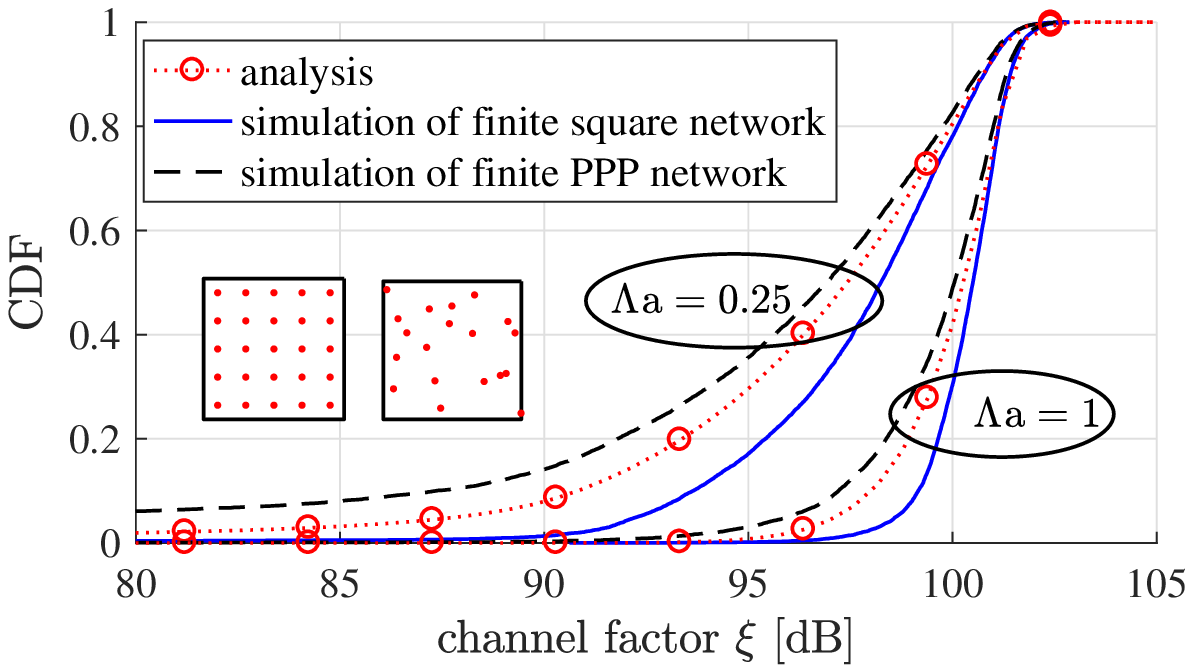}
		\caption{}
		\label{}
	\end{subfigure}	
	\begin{subfigure}{.49\textwidth}
		\centering
		\includegraphics[width=1\linewidth]{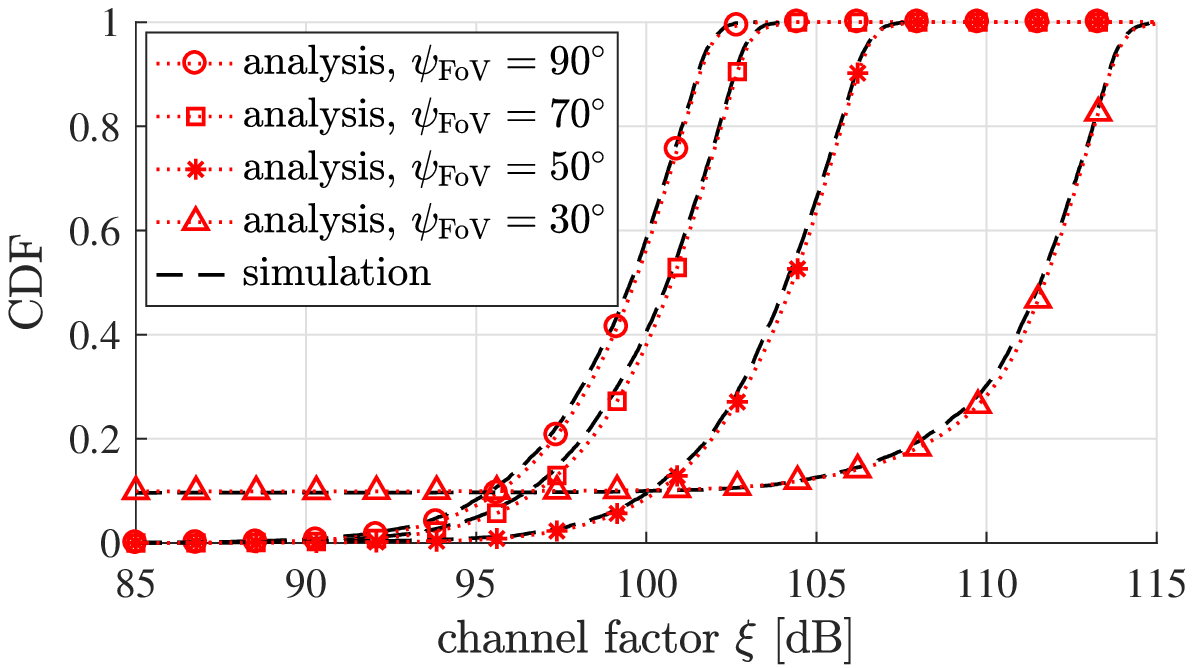}
		\caption{}
		\label{}
	\end{subfigure}%
	\caption{(a) CDF of channel factor $\xi$ with various AP density $\Lambda_{\rm a}$. The simulated channel factor statistics in a finite square network and in a PPP network have been included for comparison. The two subplots demonstrate the 2-dimensional layout of the finite square and PPP networks in the case of $\Lambda_{\rm a}=0.25$. (b) CDF of channel factor $\xi$ with various receiver FoV $\psi_{\rm FoV}$. In these numerical results, the AP density is changed to $\Lambda_{\rm a}=0.5$.}
	\label{fig:comb5}
\end{figure}

It has been found that the outage probability is strongly related to the density of the APs in the uplink system $\Lambda_{\rm a}$. In Fig.~\ref{fig:comb5}~(a), the numerical results with $\Lambda_{\rm a}=0.25$ and $\Lambda_{\rm a}=1$ are presented, which shows a significant reduction in outage probability compared to the case with $\Lambda_{\rm a}=0.1$. As discussed in Section~\ref{subsec:blockage}, the probability of blockage is higher if the horizontal separation between a UE and the corresponding associated AP gets larger. With a higher AP density $\Lambda_{\rm a}$, the average UE-AP horizontal separation gets smaller and the number of candidate APs for UE association increases, as well. Thus, the UE outage probability can be significantly decreased with an increase of AP density $\Lambda_{\rm a}$. 
In the channel factor statistical analysis, the case in a network extending infinitely in the 2-dimensional space is considered. However, a LiFi network is deployed in a room with finite space in practice, and the AP layout is likely to follow a regular pattern. Consequently, we present the simulation results considering a finite network with a room of size $\rm 10~m \times 10~m$, and both the cases with square AP layout and PPP layout are included. Similar to the observation in \cite{cbh1601}, the numerical results of the analysis are slightly worse than the simulation results in a finite square network. However, it is better than the simulation result in a finite PPP network. This is because in the boundary of a finite network, the UEs have a fewer number of candidate APs to associate with compared to the case in an infinite network. Thus, the UE has a slightly worse channel factor distribution.


The FoV of the receiver is found to be another important parameter that significantly affects the channel factor statistics. As shown in Fig.~\ref{fig:comb5}~(b), with a decrease of FoV $\psi_{\rm FoV}$, the overall channel factor level increases. This is because a smaller FoV leads to a higher concentrator gain according to \eqref{eq:G_1}, which increases the received signal power. However, a further decrease in FoV leads to a significant increase in outage probability, as shown in the case with $\psi_{\rm FoV}=30\degree$ in Fig.~\ref{fig:comb5}~(b). A decrease in FoV leads to a smaller coverage area of each AP. This is equivalent to reduce the maximum link distance of a UE. The number of APs that can connect to a UE increases with $\Lambda_{\rm a}$ as discussed above, but also decreases with the maximum link distance of a UE. Consequently, the outage probability increases when the FoV is very small.

\subsection{Average achievable data rate}
In Section~\ref{subsec:PAM_SCFDE}, we have concluded the calculation of the achievable data rate conditioning on a channel factor $\xi$ with different modulation schemes and modulation bandwidth configurations. In conjunction with the statistics of the channel factor, the average achievable data rate can be calculated as:
\begin{align}
\bar{R}_{\rm b}=\int_0^{\infty}\mathfrak{f}_{\xi}(T)R_{\rm b}(\xi)\mathrm{d}T,
\label{eq:average_data_rate}
\end{align}
where $R_{\rm b}(\xi)$ can be calculated using either \eqref{eq:data_rate_fixed} or \eqref{eq:data_rate_adaptive}. Note that equation \eqref{eq:average_data_rate} can be evaluated using numerical method. The upper limit can be replaced by a finite value $\xi_{\rm max}$ that $F_{\xi}(\xi_{\rm max})\approx 1$.

\begin{figure}
	\centering
	\begin{subfigure}{.55\textwidth}
		\centering
		\includegraphics[width=1\textwidth]{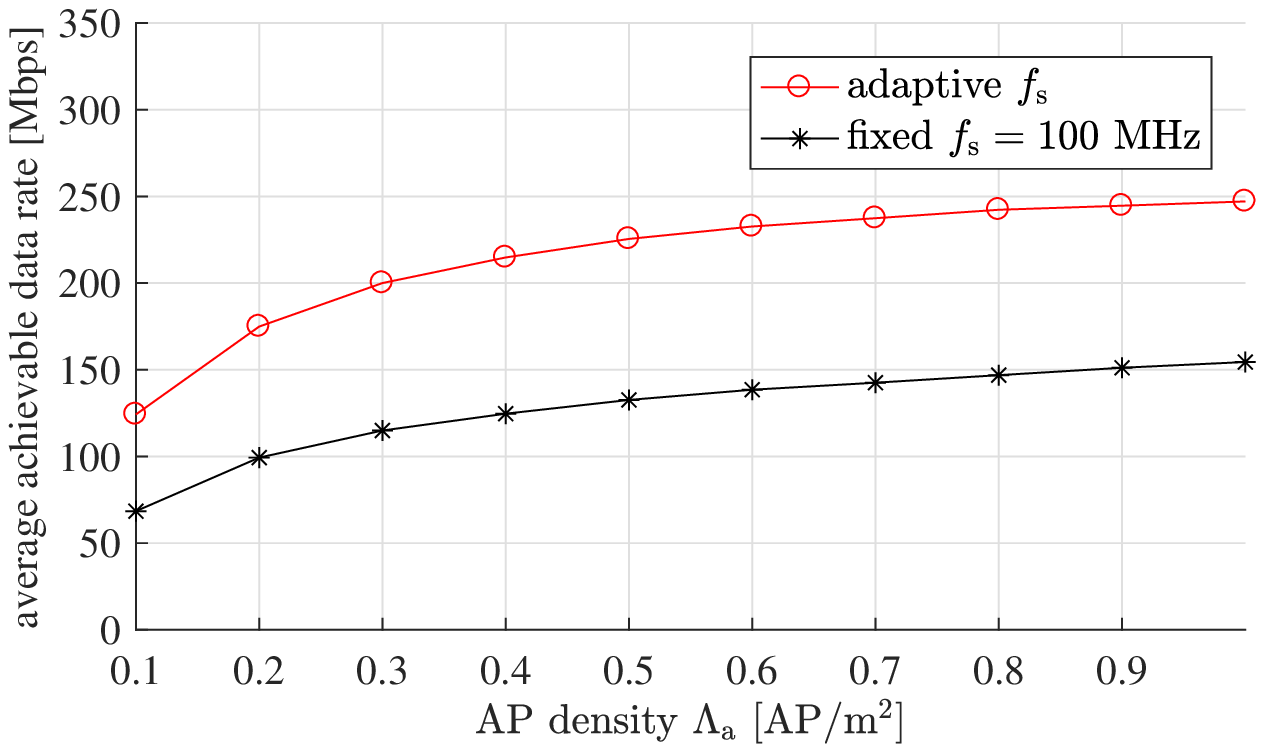}
		\caption{}
		\label{}
	\end{subfigure}	
	\begin{subfigure}{.35\textwidth}
		\centering
		\includegraphics[width=1\linewidth]{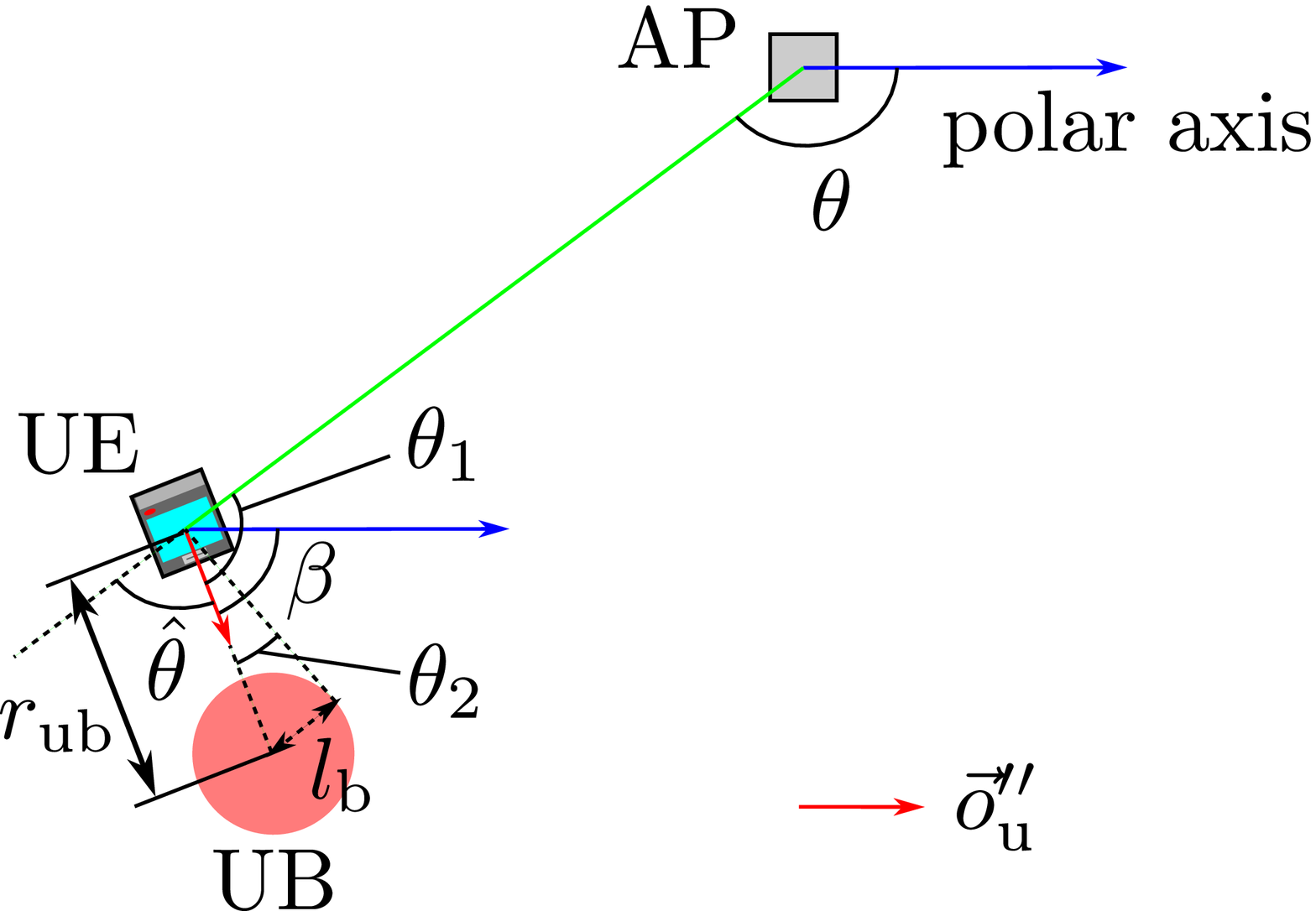}
		\caption{}
		\label{}
	\end{subfigure}%
	\caption{(a) Average achievable data rate using PAM-SCFDE with fixed and adaptive modulation bandwidth against AP density $\lambda_{\rm a}$ (b) Geometry top view of the AP, UE and UB.}
	\label{fig:comb6}
\end{figure}

The numerical results of the average achievable data rate are presented in Fig.~\ref{fig:comb6}~(a). It shows that with an increase of AP density $\Lambda_{\rm a}$, the average achievable data rate increases, as well. This is because a larger number of nearby APs can increase the probability of establishing a more reliable optical channel. In addition, it can be seen that PAM-SCFDE with adaptive modulation bandwidth is able to offer an average data rate of up to 250~Mbps. However, if the modulation bandwidth is fixed to 100~Mbps, the achievable data rate is significantly reduced.

\section{Conclusion}
\label{sec:conclusion}
In this paper, the performance of a LiFi uplink system based on wireless infrared communication was analysed, and an analytical framework was established. The achievable data rate with PAM-SCFDE was investigated under a fixed linear dynamic range constraint. In addition, a number of practical issues in the LiFi uplink, such as front-end spectral characteristics, modulation cut-off frequency, link blockage and random user device orientation were taken into account. In particular, analytical expressions for the statistics of optical channel path loss were derived. Based on these statistics, the distribution of the channel factor and the average achievable data rate were analysed. Furthermore, the configuration of AP density and receiver FoV was found to be important to the uplink performance.

\numberwithin{equation}{section}
\appendix
\subsection{Proof of Theorem~\ref{theorem:1}}
\label{app:proof of theorem 1}
The proof of Theorem~\ref{theorem:1} starts from making an assumption about the position of a UB relative to the corresponding UE. When someone is using a mobile device, he/she normally looks at the front screen of the device. Therefore, it is intuitive to assume that the UE is always directed to its UB as shown in Fig.~\ref{fig:comb6}~(b). Note that Fig.~\ref{fig:comb6}~(b) shows the top view of the uplink geometry. Regarding a UE that is located at $\vec{p}_{\rm u}$ with an orientation of $\vec{o}_{\rm u}$, the position of its UB is defined as $\vec{p}_{\rm ub}=\vec{p}_{\rm u}+r_{\rm ub}\vec{o}_{\rm u}''$, where $\vec{o}_{\rm u}''$ is the vector projection of the orientation vector $\vec{o}_{\rm u}$ on the horizontal plane with normalisation ($||\vec{o}_{\rm u}''||=1$). Recall that $r_{\rm ub}$ is the distance between the UE and its UB. In section~\ref{subsubsec:user blocker}, we have concluded that
\begin{align}
|\theta_1|\leq\theta_2~\text{and}~r\geq\frac{\hat{z}}{z_{\rm b}-z_{\rm u}}r_{\rm ub}\Rightarrow G=0.
\label{eq:self_block}
\end{align}
By combining the geometry shown in Fig.~\ref{fig:comb2}~(b) and Fig.~\ref{fig:blockage}, the relationship between $\theta$, $\beta$, $\hat{\theta}$, $\theta_1$ and $\theta_2$ can be found as that shown in Fig.~\ref{fig:comb6}~(b). Based on the geometry, we can find that $\theta_1=\pi-\hat{\theta}$, and $\theta_2=\arcsin\left(\frac{l_{\rm b}}{r_{\rm ub}}\right)$. Thus, according to \eqref{eq:self_block}, we can modify \eqref{eq:G_2} to be:
\begin{align}
G=\left\{
\begin{array}{lr}
0 : \left|\hat{\theta}-\pi\right|\leq\theta_2~\text{and}~r\geq\frac{\hat{z}}{z_{\rm b}-z_{\rm u}}r_{\rm ub}& \\
\frac{c_0\mathbf{1}_{\mathcal{V}}\left(\hat{z}\cos\alpha-r\sin\alpha\cos\hat{\theta}\right)^m}{\left(r^2+\hat{ z}^2\right)^{\frac{m+3}{2}}} : \text{otherwise}&
\end{array}\right..
\end{align}
Note that the range of $\hat{\theta}$ is $(-\pi,\pi]$. However, $G$ is an even function of $\hat{\theta}$: $G\left(\hat{\theta}\right)=G\left(-\hat{\theta}\right)$. Therefore, we can consider $\hat{\theta}$ in the range of $[0,\pi]$. Firstly, we consider the distribution of $G$ conditioning on $\alpha$. In this case, the distribution of $G$ is determined by the random variable $\hat{\theta}$. Thus, the probability that $G$ is not greater than a threshold $T$ can be interpreted as the probability that $\hat{\theta}$ is within a specified range $\Theta$: $\mathbb{P}[G\leq T|\alpha]=\mathbb{P}\left[\hat{\theta}\in\Theta\right]$, where $\Theta=\left\{\hat{\theta}\in[0,\pi]:G\left(\hat{\theta}\right)\leq T\right\}$. Next, we are going to focus on two cases: one with and one without blockage by UB. In the case that $r<\frac{\hat{z}}{z_{\rm b}-z_{\rm u}}r_{\rm ub}$, the link will not be blocked by the user. Therefore, $G$ is an increasing function of $\hat{\theta}$, and it can be found that $\Theta=\left\{0\leq\hat{\theta}\leq \theta_{\rm T}\right\}$,
where
\begin{align}
\theta_{\rm T}=\arccos\left(\frac{\hat{z}\cos\alpha-\left(T/c_0\right)^{\frac{1}{m}}\left(r^2+\hat{z}^2\right)^{\frac{3}{2m}+\frac{1}{2}}}{r\sin\alpha}\right).
\label{eq:theta_T}
\end{align}
Since $\hat{\theta}$ follows the distribution \eqref{eq:pdf_hat_theta}, the conditional probability can be found as:
\begin{align}
&\mathbb{P}[G\leq T|\alpha]=\mathbb{P}\left[\hat{\theta}\in\Theta\right]=2\int_0^{\theta_{\rm T}}\mathfrak{f}_{\hat{\theta}}\left(\hat{\theta}\right)\mathrm{d}\hat{\theta}=\frac{\theta_{\rm T}}{\pi}.
\label{eq:P_G_1}
\end{align} 
In the case that $r\geq\frac{\hat{z}}{z_{\rm b}-z_{\rm u}}r_{\rm ub}$, the link will be blocked ($G=0$) if $\hat{\theta}>\pi-\theta_2$. Based on this observation we can find that $\Theta=\left\{0\leq\hat{\theta}\leq \theta_{\rm T}\right\}\cup\left\{\pi-\theta_2\leq\hat{\theta}\leq \pi\right\}$.
The corresponding probability can be found as:
\begin{align}
&\mathbb{P}[G\leq T|\alpha]=2\int_0^{\min\left(\theta_{\rm T},\pi-\theta_2\right)}\mathfrak{f}_{\hat{\theta}}\left(\hat{\theta}\right)\mathrm{d}\hat{\theta}+2\int_{\pi-\theta_2}^{\pi}\mathfrak{f}_{\hat{\theta}}\left(\hat{\theta}\right)\mathrm{d}\hat{\theta}=\frac{\min\left(\theta_{\rm T},\pi-\theta_2\right)}{\pi}+\frac{\theta_2}{\pi}. \label{eq:P_G_2}
\end{align} 
By combining \eqref{eq:P_G_1}  with \eqref{eq:P_G_2} in conjunction with \eqref{eq:theta_T} and $\theta_2=\arcsin\left(\frac{l_{\rm b}}{r_{\rm ub}}\right)$, the conditional CDF of $G$ can be concluded as:
\begin{align}
&\mathbb{P}[G\leq T|\alpha]=\frac{\mathbf{1}_{\mathcal{S}_1}}{\pi}\arcsin\left(\frac{l_{\rm b}}{r_{\rm ub}}\right)+\frac{1}{\pi}\arccos^{\dagger}\left(\frac{\hat{z}\cos\alpha-\left(T/c_0\right)^{\frac{1}{m}}\left(r^2+\hat{z}^2\right)^{\frac{3}{2m}+\frac{1}{2}}}{r\sin\alpha}\right), 
\end{align}
where $\mathcal{S}_1=\left\{r:r\geq\frac{\hat{z}}{z_{\rm b}-z_{\rm u}}r_{\rm ub}\right\}$ and
\begin{align}
&\arccos^{\dagger}(u)=\left\{\begin{array}{lr}
\pi-\mathbf{1}_{\mathcal{S}_1}\arcsin\left(\frac{l_{\rm b}}{r_{\rm ub}}\right) : &u < -\left(1-\frac{\mathbf{1}_{\mathcal{S}_1}l_{\rm b}^2}{r_{\rm ub}^2}\right)^{\frac{1}{2}} \\
\arccos(u): &-\left(1-\frac{\mathbf{1}_{\mathcal{S}_1}l_{\rm b}^2}{r_{\rm ub}^2}\right)^{\frac{1}{2}} \leq u \leq 1  \\
0 : &u>1 
\end{array}\right..
\label{eq:arccos_dagger} 
\end{align}
Then we consider the distribution of the channel path loss with random $\alpha$ following the distribution \eqref{eq:pdf_alpha}. The CDF of $G$ can be calculated using: $\mathbb{P}[G\leq T]=\int_{0}^{\frac{\pi}{2}} \mathfrak{f}_{\alpha}(\alpha)\mathbb{P}[G\leq T|\alpha]\mathrm{d}\alpha$.
Since $\mathfrak{f}_{\alpha}(\alpha)\approx 0$ when $\alpha>\pi/2$ or $\alpha<0$, we can relax the integration limits to the range of $(+\infty,-\infty)$:
\begin{align}
&\mathbb{P}[G\leq T]=\int\limits_{-\infty}^{\infty}\mathfrak{f}_{\alpha}(\alpha)\mathbb{P}[G\leq T|\alpha]\mathrm{d}\alpha=\int\limits_{-\infty}^{\infty}\frac{\mathrm{e}^{-\frac{|\alpha-\mu_{\rm L}|}{b_{\rm L}}}\arccos^{\dagger}\left(\mathcal{Y}\right)}{2b_{\rm L}\pi}\mathrm{d}\alpha+\frac{\mathbf{1}_{\mathcal{S}_1}\arcsin\left(\frac{l_{\rm b}}{r_{\rm ub}}\right)}{\pi}, \label{eq:CDF_G_1}
\end{align}
where
\begin{align}
\mathcal{Y}=\frac{\hat{z}\cos\alpha-\left(T/c_0\right)^{\frac{1}{m}}\left(r^2+\hat{z}^2\right)^{\frac{3}{2m}+\frac{1}{2}}}{r\sin\alpha}. \label{eq:Y_1}
\end{align}
Note that \eqref{eq:arccos_dagger} is a piecewise function. In order to solve the integration \eqref{eq:CDF_G_1}, we have to find the range of $\alpha$ that matches the function \eqref{eq:arccos_dagger} in various cases. By solving the equations $\mathcal{Y}=1$ and $\mathcal{Y}=-\sqrt{1-\mathbf{1}_{\mathcal{S}_1}l_{\rm b}^2/r_{\rm ub}^2}$, the following boundary values for $\alpha$ can be found:
\begin{align}
	&{\alpha}_{\rm min}=\left|\arcsin\left[\frac{\left(T/c_0\right)^{\frac{1}{m}}\left(r^2+\hat{z}^2\right)^{\frac{3}{2m}}}{ \left(1-\frac{\mathbf{1}_{\mathcal{S}_1}\mathbf{1}_{\mathcal{S}_2}r^2l_{\rm b}^2}{r_{\rm ub}^2\left(r^2+\hat{z}^2\right)}\right)^{\frac{1}{2}}}\right]-\arctan\left[\frac{\hat{z}}{r}\left(1-\frac{\mathbf{1}_{\mathcal{S}_1}\mathbf{1}_{\mathcal{S}_2}l_{\rm b}^2}{r_{\rm ub}^2}\right)^{-\frac{1}{2}}\right]\right|, \label{eq:alpha_min} \\
	&{\alpha}_{\rm max}=\pi-\arcsin\left[\frac{\left(T/c_0\right)^{\frac{1}{m}}\left(r^2+\hat{z}^2\right)^{\frac{3}{2m}}}{ \left(1-\frac{\mathbf{1}_{\mathcal{S}_1}r^2l_{\rm b}^2}{r_{\rm ub}^2\left(r^2+\hat{z}^2\right)}\right)^{\frac{1}{2}}}\right]-\arctan\left[\frac{\hat{z}}{r}\left(1-\frac{\mathbf{1}_{\mathcal{S}_1}l_{\rm b}^2}{r_{\rm ub}^2}\right)^{-\frac{1}{2}}\right], \label{eq:alpha_max}
\end{align}
where $\mathcal{S}_2=\left\{T:T>\mathcal{G}_0=c_0\hat{z}^m\left(r^2+\hat{z}^2\right)^{-\frac{m+3}{2}}\right\}$. The function $\arcsin(u)$ is modified to have a domain of $(+\infty,-\infty)$, where $\arcsin(u)=\pi/2$ with $u>1$ and $\arcsin(u)=-\pi/2$ with $u<-1$.
With the $\alpha$ boundary values \eqref{eq:alpha_min} and \eqref{eq:alpha_max}, we can decompose the integration \eqref{eq:CDF_G_1} as:
\begin{align}
&\mathbb{P}[G\leq T]=\frac{\mathbf{1}_{\mathcal{S}_1}}{\pi}\arcsin\left(\frac{l_{\rm b}}{r_{\rm ub}}\right)+ \int_{-\infty}^{\alpha_{\rm min}}\frac{\mathrm{e}^{-\frac{|\alpha-\mu_{\rm L}|}{b_{\rm L}}}}{2b_{\rm L}\pi}\arccos^{\dagger}\left(\mathcal{Y}\right)\mathrm{d}\alpha \nonumber \\
&+ \int_{\alpha_{\rm max}}^{+\infty}\frac{\mathrm{e}^{-\frac{|\alpha-\mu_{\rm L}|}{b_{\rm L}}}}{2b_{\rm L}\pi}\arccos^{\dagger}\left(\mathcal{Y}\right)\mathrm{d}\alpha+ \int_{\alpha_{\rm min}}^{\alpha_{\rm max}}\frac{\mathrm{e}^{-\frac{|\alpha-\mu_{\rm L}|}{b_{\rm L}}}}{2b_{\rm L}\pi}\arccos^{\dagger}\left(\mathcal{Y}\right)\mathrm{d}\alpha. \label{eq:PGT}
\end{align}
Now we consider the first integral in \eqref{eq:PGT}. It has been found that under the condition $\alpha\leq\alpha_{\rm min}$,
\begin{align}
\arccos^{\dagger}(\mathcal{Y})=\left\{\begin{array}{lr}
\pi-\mathbf{1}_{\mathcal{S}_1}\arcsin\left(\frac{l_{\rm b}}{r_{\rm ub}}\right) : &T>\mathcal{G}_0  \\
0 : &T\leq\mathcal{G}_0 
\end{array}\right..
\label{eq:accos_Y_1} 
\end{align}
Inserting \eqref{eq:accos_Y_1} into the first integral in \eqref{eq:PGT} leads to:
\begin{align}
&\int_{-\infty}^{\alpha_{\rm min}}\frac{\mathrm{e}^{-\frac{|\alpha-\mu_{\rm L}|}{b_{\rm L}}}}{2b_{\rm L}\pi}\arccos^{\dagger}\left(\mathcal{Y}\right)\mathrm{d}\alpha \nonumber \\ &=\frac{\mathbf{1}_{\mathcal{S}_2}}{2}\left[1+\mathrm{sgn}(\alpha_{\rm min}-\mu_{\rm L})\left(1-\mathrm{e}^{-\frac{|\alpha_{\rm min}-\mu_{\rm L}|}{b_{\rm L}}}\right)\right]\left[1-\frac{\mathbf{1}_{\mathcal{S}_1}}{\pi}\arcsin\left(\frac{l_{\rm b}}{r_{\rm ub}}\right)\right].
\label{eq:int1}
\end{align}
The calculation of the second integral in \eqref{eq:PGT} is similar. Under the condition $\alpha\geq\alpha_{\rm max}$, it has been found that 
$\arccos^{\dagger}\left(\mathcal{Y}\right)=\pi-\mathbf{1}_{\mathcal{S}_1}\arcsin\left(l_{\rm b}/r_{\rm ub}\right)$. Therefore, the second integral in \eqref{eq:PGT} can be derived as:
\begin{align}
&\int_{\alpha_{\rm max}}^{+\infty}\frac{\mathrm{e}^{-\frac{|\alpha-\mu_{\rm L}|}{b_{\rm L}}}}{2b_{\rm L}\pi}\arccos^{\dagger}\left(\mathcal{Y}\right)\mathrm{d}\alpha \nonumber \\ &=\frac{1}{2}\left[1-\mathrm{sgn}(\alpha_{\rm max}-\mu_{\rm L})\left(1-\mathrm{e}^{-\frac{|\alpha_{\rm max}-\mu_{\rm L}|}{b_{\rm L}}}\right)\right]\left[1-\frac{\mathbf{1}_{\mathcal{S}_1}}{\pi}\arcsin\left(\frac{l_{\rm b}}{r_{\rm ub}}\right)\right].
\label{eq:int2}
\end{align}
The third integral in \eqref{eq:PGT} is hard to solve directly. However, it can be estimated with the approximations stated in Section~\ref{sec:channel_path_loss_statistics}. According to the Taylor series formula, we define the following approximation: 
\begin{align}
\arccos(\mathcal{Y})\approx\arccos(\mathcal{Y}_0)-\frac{\mathcal{Y}-\mathcal{Y}_0}{\sqrt{1-\mathcal{Y}_0^2}}.
\label{eq:acos_Y_approx}
\end{align}
The expansion is about a specific point, $\mathcal{Y}_0$, which has a value in the range of $[-1,1]$. It has been found that by assigning different values to $\mathcal{Y}_0$ for various threshold value $T$, the approximation accuracy can be significantly improved. However, we will address the configuration of $\mathcal{Y}_0$ in the later part of this proof. It is noted that the random elevation angle $\alpha$ varies about its mean value $\mu_{\rm L}$. The probability that the value of $\alpha$ is very close to $\mu_{\rm L}$ is higher than the probability that $\alpha$ is significantly deviated from $\mu_{\rm L}$. Thus, we define a new random angle $\tilde{\alpha}=\alpha-\mu_{\rm L}$ with zero mean. Considering the approximation $\sin\tilde{\alpha}=\tilde{\alpha}$ and $\cos\tilde{\alpha}=1$ for small $\tilde{\alpha}$, the trigonometric terms in \eqref{eq:Y_1} can be approximated by $\cos\alpha\approx\cos\mu_{\rm L}-\tilde{\alpha}\sin\mu_{\rm L}$ and $\sin\alpha\approx\sin\mu_{\rm L}+\tilde{\alpha}\cos\mu_{\rm L}$. Thus, \eqref{eq:Y_1} can be rewritten as:
\begin{align}
\mathcal{Y}&\approx\frac{\hat{z} \left(1-\tilde{\alpha}\tan\mu_{\rm L}\right)-\left(T/c_0\right)^{\frac{1}{m}}\left(r^2+\hat{z}^2\right)^{\frac{3}{2m}+\frac{1}{2}}\cos^{-1}\mu_{\rm L}}{r\left(\tan\mu_{\rm L}+\tilde{\alpha}\right)}.
\label{eq:Y_approx}
\end{align}
Under the condition $\alpha_{\rm min}<\alpha<\alpha_{\rm max}$, it is found that $\arccos^{\dagger}\left(\mathcal{Y}\right)=\arccos\left(\mathcal{Y}\right)$. In conjunction with \eqref{eq:acos_Y_approx} and \eqref{eq:Y_approx}, the third integral in \eqref{eq:PGT} can be rewritten as:
\begin{align}
&\int_{\alpha_{\rm min}}^{\alpha_{\rm max}}\frac{\mathrm{e}^{-\frac{|\alpha-\mu_{\rm L}|}{b_{\rm L}}}}{2b_{\rm L}\pi}\arccos^{\dagger}\left(\mathcal{Y}\right)\mathrm{d}\alpha\approx\left(\frac{\arccos(\mathcal{Y}_0)}{2 \pi b_{\rm L}}+\frac{\mathcal{Y}_0+\frac{\hat{z}}{r}\tan\mu_{\rm L}}{2 \pi b_{\rm L}\sqrt{1-\mathcal{Y}_0^2}}\right)\int_{\tilde{\alpha}_{\rm min}}^{\tilde{\alpha}_{\rm max}}\mathrm{e}^{-\frac{|\tilde{\alpha}|}{b_{\rm L}}}\mathrm{d}\tilde{\alpha}\nonumber \\ &-\frac{\hat{z}\cos^{-2}\mu_{\rm L}\frac{}{}-\left(T/c_0\right)^{\frac{1}{m}}\left(r^2+\hat{z}^2\right)^{\frac{3}{2m}+\frac{1}{2}}\cos^{-1}\mu_{\rm L}}{2 \pi b_{\rm L}r\sqrt{1-\mathcal{Y}_0^2}}\int_{\tilde{\alpha}_{\rm min}}^{\tilde{\alpha}_{\rm max}}\frac{\mathrm{e}^{-\frac{|\tilde{\alpha}|}{b_{\rm L}}}}{\tan\mu_{\rm L}+\tilde{\alpha}}\mathrm{d}\tilde{\alpha},
\label{eq:int3}
\end{align}
where $\tilde{\alpha}_{\rm min}={\alpha}_{\rm min}-\mu_{\rm L}$, $\tilde{\alpha}_{\rm max}={\alpha}_{\rm max}-\mu_{\rm L}$ and the integral in \eqref{eq:int3} can be calculated by:
\begin{align}
&\int_{\tilde{\alpha}_{\rm min}}^{\tilde{\alpha}_{\rm max}}\mathrm{e}^{-\frac{|\tilde{\alpha}|}{b_{\rm L}}}\mathrm{d}\tilde{\alpha}=b_{\rm L}\left(\mathrm{e}^{\frac{[\tilde{\alpha}_{\rm max}]^{-}}{b_{\rm L}}}-\mathrm{e}^{\frac{[\tilde{\alpha}_{\rm min}]^{-}}{b_{\rm L}}}-\mathrm{e}^{-\frac{[\tilde{\alpha}_{\rm max}]^{+}}{b_{\rm L}}}+\mathrm{e}^{-\frac{[\tilde{\alpha}_{\rm min}]^{+}}{b_{\rm L}}}\right) \label{eq:int_e1} \\
&\int_{\tilde{\alpha}_{\rm min}}^{\tilde{\alpha}_{\rm max}}\frac{\mathrm{e}^{-\frac{|\tilde{\alpha}|}{b_{\rm L}}}}{\tan\mu_{\rm L}+\tilde{\alpha}}\mathrm{d}\tilde{\alpha}=\mathrm{e}^{-\frac{\tan\mu_{\rm L}}{b_{\rm L}}}\left(\mathrm{Ei}\left[\frac{[\tilde{\alpha}_{\rm max}]^{-}+\tan\mu_{\rm L}}{b_{\rm L}}\right]-\mathrm{Ei}\left[\frac{[\tilde{\alpha}_{\rm min}]^{-}+\tan\mu_{\rm L}}{b_{\rm L}}\right]\right) \nonumber \\ &+\mathrm{e}^{\frac{\tan\mu_{\rm L}}{b_{\rm L}}}\left(\mathrm{Ei}\left[-\frac{[\tilde{\alpha}_{\rm max}]^{+}+\tan\mu_{\rm L}}{b_{\rm L}}\right]-\mathrm{Ei}\left[-\frac{[\tilde{\alpha}_{\rm min}]^{+}+\tan\mu_{\rm L}}{b_{\rm L}}\right]\right). \label{eq:int_e2}
\end{align}
By inserting \eqref{eq:int1}, \eqref{eq:int2} and \eqref{eq:int3} into \eqref{eq:PGT}, the calculation of $\mathbb{P}[G\leq T]$ with approximation can be obtained, which is denoted by $\mathbb{P}_{\rm approx}[G\leq T]$.

Now we consider the configuration of $\mathcal{Y}_0$. Firstly, we define the approximation error as $e=\mathbb{P}_{\rm approx}[G\leq T]-\mathbb{P}[G\leq T]$. Then, we evaluate the square of the approximation error $|e|^2$ with all possible values of $\mathcal{Y}_0$ within the range of $[-1,1]$ for a given threshold $T$. Note the $\mathbb{P}[G\leq T]$ is calculated using \eqref{eq:PGT} with a numerical integration. The value of $|e|^2$ is compared with a threshold $e_{\rm T}$ which defines a maximum acceptable approximation error. $e_{\rm T}=1\times10^{-2}$ has been used in our test. Next, the region of $\mathcal{Y}_0$ that lead to acceptable approximation accuracy can be concluded as: $\Upsilon=\{\mathcal{Y}_0\in[-1,1]:|e|^2\leq e_{\rm T}\}$. Finally, the function for $\mathcal{Y}_0$ of $T$ can be empirically designed so that $\mathcal{Y}_0\in\Upsilon$ for $T\in[0,\mathcal{G}_{\rm max}]$ and other parameters in the region of interest. The following function is one of the candidates, which offers high approximation accuracy and relatively low complexity.
\begin{align}
\mathcal{Y}_0=\left\{\begin{array}{lr} -0.5 &: 0\leq T\leq \mathcal{G}_1 \\ \frac{\left(\zeta(r)+0.5\right)(T-\mathcal{G}_1)}{\mathcal{G}_2-\mathcal{G}_1}-0.5 &: \mathcal{G}_1<T<\mathcal{G}_{\rm th} \\ 0 &: T\geq \mathcal{G}_{\rm th}~\text{or}~T<0 \end{array} \right.,
\label{eq:Y_0_example}
\end{align}
where the related parameters can be found in Table~\ref{table:minor_expressions}. Note that the calculation of $\mathbb{P}_{\rm approx}[G\leq T]$ is not limited to \eqref{eq:Y_0_example}. Other candidates can also be found as long as the function fulfils the condition $\mathcal{Y}_0\in\Upsilon$.

Now we include the effect of blockage caused by NUBs on the optical channel. Since this type of blockage is independent of the status of the optical channel path loss, the analytical result \eqref{eq:NUB_P_b} presented in Section~\ref{subsubsec:nonuser_blockage} can be directly used to calculate the final CDF of the optical channel:
\begin{align}
F_G(T)=\bar{\mathcal{P}}_{\rm b}+\left(1-\bar{\mathcal{P}}_{\rm b}\right)\mathbb{P}_{\rm approx}[G\leq T].
\label{eq:CDF_G_all_1}
\end{align}
By combining the analytical expressions \eqref{eq:int1}, \eqref{eq:int2}, \eqref{eq:int3}, \eqref{eq:PGT}, \eqref{eq:Y_0_example} and \eqref{eq:CDF_G_all_1} the CDF result of the optical channel path loss shown in \eqref{eq:CDF_G_r} can be obtained.


%

\ifCLASSOPTIONcaptionsoff
  \newpage
\fi



%
%
%
%

\bibliographystyle{IEEEtranTCOM}
\bibliography{cheng_collect}

\end{document}